\documentclass{scrartcl}
\usepackage[utf8]{inputenc}
\usepackage[natbibapa]{apacite}
\usepackage{graphicx}
\usepackage[hidelinks]{hyperref}
\usepackage{color}
\usepackage{amsmath}
\usepackage{authblk}
\usepackage[section]{placeins}
\usepackage[position=top]{subfig}
\usepackage{rotating}
\usepackage{tabularx}
\usepackage{ltablex}

\setkomafont{disposition}{\normalcolor\bfseries}

\newcommand\blfootnote[1]{%
  \begingroup
  \renewcommand\thefootnote{}\footnote{#1}%
  \addtocounter{footnote}{-1}%
  \endgroup
}

\title{Fairness in Algorithmic Profiling: A German Case Study}

\author[1,*]{Christoph Kern}
\author[1,*]{Ruben L. Bach}
\author[2]{Hannah Mautner}
\author[3,4]{Frauke Kreuter}

\affil[*]{Co-first authors}
\affil[1]{School of Social Sciences, University of Mannheim, Germany}
\affil[2]{dmTECH, Karlsruhe, Germany}
\affil[3]{Department of Statistics, LMU Munich, Germany}
\affil[4]{Joint Program in Survey Methodology, University of Maryland, USA}

\date{August 2021}

\begin{document}

\maketitle

\begin{abstract}
Algorithmic profiling is increasingly used in the public sector as a means to allocate limited public resources effectively and objectively. One example is the prediction-based statistical profiling of job seekers to guide the allocation of support measures by public employment services. However, empirical evaluations of potential side-effects such as unintended discrimination and fairness concerns are rare. In this study, we compare and evaluate statistical models for predicting job seekers' risk of becoming long-term unemployed with respect to prediction performance, fairness metrics, and vulnerabilities to data analysis decisions. Focusing on Germany as a use case, we evaluate profiling models under realistic conditions by utilizing administrative data on job seekers' employment histories that are routinely collected by German public employment services. Besides showing that these data can be used to predict long-term unemployment with competitive levels of accuracy, we highlight that different classification policies have very different fairness implications. We therefore call for rigorous auditing processes before such models are put to practice.\blfootnote{\textbf{Acknowledgements:} Christoph Kern's and Ruben Bach's work was supported by the Baden-W{\"u}rttemberg Stiftung (grant “FairADM -- Fairness in Algorithmic Decision Making” to Ruben Bach, Christoph Kern and Frauke Kreuter). Hannah Mautner worked on the project while she was a graduate student at the Institute for Employment Research (IAB) in Nuremberg, Germany.} 
\end{abstract}

\section{Introduction}
\label{Intro}

Statistical profiling presents an increasingly important avenue for informing high-stakes policy decisions such as the allocation of scarce public resources in a variety of settings. Examples include the allocation of intervention and supervision resources in criminal justice \citep{howard2012construction}, the allocation of in-person investigations in the context of child protection services \citep{chouldechova2018case}, and the allocation of home inspections to identify and control health hazards \citep{Potash2020}. In these scenarios, statistical models are used to provide an initial risk assessment such that decisions can be made, for example, regarding the question which cases require special attention and should be prioritized. So far, risk assessments in the government sector are often still based on human experience or on a (small) set of pre-defined rules. The hope is, however, that statistical and algorithmic profiling will increase both effectiveness and objectivity of the decision-making process. For example, psychological studies of decision-making have demonstrated for more than 60 years that statistical models and simple rule-based predictions outperform humans in tasks such as predicting academic success, job performance and psychiatric prognosis \citep{meehl1954clinical, yu2020pushing, kahneman2021noise}. That is, a statistical model is often simply more accurate in predicting, for example, who is likely to recidivate, and such models will do so consistently by returning similar outputs for similar instances. When compared to predefined rules, statistical models commonly offer an additional performance advantage \citep{Helsby2018, Pan2017} that is fueled by rapid advances in the field of machine learning and the increasing availability of various large data sources.  

However, implementing an algorithmic profiling system in practice involves a number of critical design decisions. Questions that need to be answered include, for example, what type of prediction method should be applied? Which type of information should be used for model training? How should resources be allocated based on a prediction model's outputs? Eventually, such decisions can substantially affect the extent to which different societal groups are targeted or reached by support programs and public services. This especially includes the risk of perpetuating discrimination against historically disadvantaged groups: A profiling model that was built to classify job seekers in Austria, for example, exhibited a negative effect of being female on short-term re-employment propensities, that is, the probability of finding employment for at least 3 months within the next 7 months \citep{holl2018}. Based on such a model, different classification policies could be applied under which female job seekers could have higher (prioritize job seekers with low predicted re-employment propensities) or lower (prioritize job seekers with medium predicted re-employment propensities) chances of receiving extensive support by employment agencies, compared to their male counterparts. The same model, however, did show relatively high accuracy scores (80\% for female and male, \citealt{holl2018}) and thus could improve over human ad-hoc profiling with respect to the ability of identifying job seekers with low re-integration chances. Similar concerns have been voiced regarding an (in)famous statistical prediction tool used in the U.S. criminal justice system \citep{angwin2016machine}.

Against this background, we compare and evaluate statistical models for predicting job seekers' risk of becoming long-term unemployed (LTU) with respect to prediction performance, fairness metrics, and vulnerabilities to data analysis decisions in this study. Focusing on Germany as a use case, we evaluate profiling models by utilizing administrative data on job seekers' employment histories that are routinely collected by German public employment services. Our contribution to the literature on algorithmic profiling and fairness in profiling is twofold: (1) We report on the performance of different prediction methods and thus on the potentials of implementing algorithmic profiling of job seekers in Germany under realistic conditions. (2) We evaluate fairness implications of data analysis decisions such as using different classification thresholds and training data histories. This analysis particularly focuses on the socio-structural composition of the groups that would eventually be prioritized and targeted under different policies. 

We use regression and machine learning techniques, specifically, logistic regression, penalized logistic regression, random forests and gradient boosting to build profiling models. For each technique, we train two sets of prediction models that differ in the time frame that is used for model training. For each model, three classification policies for prioritizing job seekers are implemented that focus on very high (top 10\%), high (top 25\%) and medium (middle 50\% of risk scores) predicted risks of LTU. Next to comparing the profiling models with respect to prediction performance, we study fairness implications of the models' classifications based on statistical parity difference, conditional statistical parity difference and consistency. We focus on four groups of job seekers: Female, non-German, female non-German and male non-German individuals. A large body of literature shows evidence of discrimination on the labor market with respect to gender \citep{Bishu2017} and ethnicity \citep{Zschirnt2016}, which is likely to be reflected in historical labor market records and thus may be learned by a prediction model. Our fairness evaluation therefore aims to study whether discrimination against these groups could be perpetuated or mitigated under a given algorithmic profiling scenario.

This paper is structured as follows: Section \ref{Background} provides an introduction to statistical profiling for prioritizing job seekers (\ref{Profiling}) and to its applications in various countries (Section \ref{ProfilingOECD}), summarizes current practices in Germany (Section \ref{ProfilingGermany}) and discusses fairness concerns in the context of algorithmic profiling (Section \ref{FairProfiling}). Section \ref{Methods} presents the data (Section \ref{Data}) and the prediction setup (Section \ref{Setup}) that is used for predicting LTU in our empirical application. The results are summarized in  Section \ref{Results}, which includes the evaluation of prediction performance (Section \ref{Performance}) and fairness measures (Section \ref{Fairness}). We discuss our findings in  Section \ref{Discussion} and provide a summary of limitations and potential next steps in  Section \ref{Limitations}. 

\section{Background}
\label{Background}

\subsection{Profiling techniques to prevent long-term unemployment}
\label{Profiling}

Fighting long-term unemployment (LTU; unemployment that lasts for more than one year) is a major societal challenge in many countries \citep{duell2016long}. LTU has serious consequences for affected individuals, not only in terms of economic deprivation but also regarding their physical and mental health as well as their overall well-being. From a societal perspective, LTU is associated with high costs for health care systems and welfare services.

Public employment services (PES) in many countries use a variety of profiling techniques to fight LTU by \textit{preventing} it through identifying those at risk of becoming long-term unemployed at an early stage \citep{loxha2014profiling}. Profiling techniques are usually designed to segment individuals at entry into unemployment into groups with similar risks of resuming work. Public employment services can then \textit{target} specific groups of individuals with treatments such as active labor market policies (measures aimed at increasing an individual's chance of finding a new job such as further vocational training and short classroom training, hiring subsidies for employers and job creation schemes). For example, those identified as having a low probability of resuming work could be given more extensive support than those with a high probability. Other PES may employ strategies that minimize overall LTU by targeting e.g., those who will profit the most. 

Three types of profiling for labor market policy can be distinguished: case worker-based profiling, rule-based profiling and statistical profiling \citep{loxha2014profiling}. While case worker-based profiling heavily relies on the expertise of the case worker in determining if, when and how a job seeker should be supported in finding a new job, rule-based profiling usually applies predefined rules such as passing a minimum time-in-unemployment threshold and certain demographic attributes to determine eligibility and the type of public employment service support. Statistical profiling, however, emphasizes the role of statistical models in identifying those who need support and in selecting the optimal treatment. For example, using historical labor market records, a prediction model could be trained to identify those who have the highest risk to become long-term unemployed. Likewise, an algorithm could be used to select the optimal treatment for a given individual.

\subsection{Statistical profiling of the unemployed across the globe}
\label{ProfilingOECD}

Several OECD countries developed a variety of statistical profiling techniques with respect to estimating chances of reintegration of the unemployed into the labor market. Unfortunately, detailed documentation of the statistical models is often not available \citep[for an exception, see][]{holl2018}. Comprehensive reviews of approaches are presented in \citet{loxha2014profiling} and \citet{Desiere2019}. Here, we will briefly review some examples from the literature.

Australia was one of the first countries that adopted data-driven profiling for labor market policy in the 1990s. Through its \textit{Job Seeker Classification Instrument}, unemployed individuals seeking a new job are classified into risk categories that describe an individual's relative level of labor market disadvantage. The classification is then used to determine the eligibility for assistance by employment services in seeking a new employment \citep{mcdonald2003risk, loxha2014profiling, caswell2010unemployed}. A score derived from 18 weighted predictors, including socio-demographic information such as age, gender, and education as well as measures capturing an individual's (un)employment history, is used to determine the classification of an individual. Weights for the individual predictors are obtained from a logistic regression model that has been updated multiple times since its first introduction \citep{lipp2005}.

Statistical profiling for labor market policy has also a long tradition in the U.S. under the Worker Profiling and Reemployment Services (WPRS) initiative \citep{Desiere2019}. States have been mandated since the 1990s to use statistical profiling to predict which unemployment-benefit claimants will exhaust their entitlement period. Those predicted to exhaust their benefit claims are mandated to attend reemployment workshops and information on outcomes are monitored subsequently to monitor benefit eligibility \citep{Pope2011, oconnell2009}.  Prediction is usually based on logistic regression models. The number of predictors varies by state but common predictors include information such as regional unemployment rate, individual occupation history and education.

Statistical profiling is also discussed, tested, or used in countries such as Austria \citep{holl2018}, Belgium \citep{desiere2021}, Denmark \citep{caswell2010unemployed}, Finland \citep{Viljanen2020}, Ireland \citep{oconnell2009, o2012transition}, the Netherlands \citep{wijnhoven}, New Zealand \citep{Desiere2019}, Poland \citep{niklas_profiling_2015}, Portugal \citep{de2018predicting} and Sweden \citep{arbetssweden}. While all rely on statistical prediction approaches to segment users, the outcomes to be predicted vary. Some models are trained to predict LTU (e.g. Belgium, Denmark, and the Netherlands), others are trained to predict exit into employment (e.g., Ireland). Regarding the statistical methods used, most of the profiling approaches are based on logistic regression models (e.g., Austria, Italy, Netherlands, Sweden) but systems based on machine learning algorithms such as random forests and gradient boosting are also used in some countries \citep[e.g., Belgium and New Zealand,][]{Desiere2019}. Most of the approaches are based on administrative labor market data like those used in this paper, but information collected from interviews conducted when newly unemployed individuals register with their PES are also used in some countries. In Belgium, data from interactions with the website of the PES, such as clicking on job vacancies posted on the website, are included as well \citep{desiere2021}. Regarding prediction performance, accuracy values from 60\% to 86\% and ROC-AUC scores between .63 and .83 are reported \citep{Desiere2019}. Prediction performance does not seem to vary much between different statistical prediction models, however \citep{matty2013, Desiere2019}. Compared to human predictions of LTU, statistical models achieve much higher prediction performance, however \citep{Arni2015,arbetssweden}.

Predicting the risk of LTU or similar outcomes (see above) is only the first step in developing a statistical profiling system. The second step is implementing a decision routine based on predicted probabilities. Due to differences in labor market policy and legislative frameworks, there is considerable variation regarding the question which risk groups are targeted by PES, based on their estimated risk scores. Many countries, however, appear to target unemployed individuals with a high LTU  risk \citep{Desiere2019}.

\subsection{Profiling practices of the unemployed in Germany}
\label{ProfilingGermany}

In Germany, profiling for supporting the unemployed was introduced as a major strategy in labor market policy during the course of the fundamental labor market reforms of 2002--2005. Triggered by high unemployment rates paired with a high share of long-time unemployment (up to 50\% in the early 2000s), the timely activation of unemployed welfare recipients with the goal of re-integrating them into the labor market to avoid LTU was identified as a main objective of the reforms \citep{bernhard2014courses}. As a consequence, the public employment service of Germany, the Federal Employment Agency, developed a strategy to combat long-term unemployment by following recommendations of the OECD regarding the minimization of the number of unemployed individuals who drift into long periods of unemployment. The strategy relies on case worker-based profiling of the unemployed and emphasizes special support for those with a high risk of entering LTU. Using a four-phase model, case workers estimate job seekers' reemployment chances ("Marktnähe") and the time it will take them to obtain a new employment. 

Before 2017, job-seekers were placed into one of six client profiles based on the estimated unemployment duration \citep{oschmiansky2020,BA2012}. Three of them comprised job seekers with estimated unemployment duration of more than 12 months and special support was given to those job seekers. In 2016/2017, placement into the profiles was removed, while the general idea of estimating job seekers reemployment chances and estimating their unemployment duration remained \citep{BA2016}. So far, statistical profiling is used to a very small extent only. Integrating a more extensive data-driven profiling has been discussed by the agency since the early 2000's, however \citep{rudolph2001profiling}. 

\subsection{Fairness concerns in statistical profiling}
\label{FairProfiling}

Many hope that using statistical profiling in public policy decisions will enhance overall government efficiency and public service delivery \citep{lepri2018}. By removing human judgment, statistical profiling may be consistent, neutral and objective and can eliminate biases and errors that humans may make. However, training statistical prediction models for profiling purposes on incomplete or biased historical data may also quickly result in discrimination towards minority groups or otherwise disadvantaged groups \citep{barocas2016, lepri2018,mehrabi_survey_2019}.

\citet{angwin2016machine}, for example, document in a widely cited study that COMPAS, a profiling system popular in the U.S. criminal justice system, systematically discriminates against black defendants. COMPAS is based on a model that predicts a defendant’s likelihood to be rearrested for a new crime while awaiting trial for the first crime. Briefly speaking, defendants with a low score are recommended for bail while those with a high score are recommended to be detained. The algorithm underlying the COMPAS system must not include race as a predictor due to U.S. anti-discrimination legislation. Still, \citet{angwin2016machine} found that black defendants were far more likely than white defendants to be incorrectly assessed high risk, while white defendants were more likely than black defendants to be incorrectly labeled as low risk. The reason why the profiling tool assigns blacks higher scores than whites is likely the disproportionately high number of blacks in jail, often for minor crimes. Since the profiling system is trained with historical data, historical discrimination against blacks will also be picked up by the profiling tool.

Similar concerns of discrimination are reported for statistical profiling of the unemployed \citep{allhutter_algorithmic_2020, desiere2021}. Because women and individuals with a migration background are disproportionately affected by unemployment and have lower job prospects \citep[e.g.,][]{kogan2011new,azmat2006gender, arntz2009unemployment, jacob2014marriage}, a statistical prediction model will quickly pick up and incorporate such associations in its predictions. Moreover, even if such characteristics are not explicitly used for training a profiling system, as in the COMPAS example above, predictions could nonetheless be affected. If labor market histories of, for example, women and men are distinct, then it is likely that a statistical algorithm will learn different patterns for women and men based on the correlation of gender and labor market histories.

Fairness of statistical profiling systems of the unemployed has not been discussed much until recently. \citet{allhutter_algorithmic_2020} study fairness concerns in the Austrian statistical profiling tool \textit{AMS algorithm}. This tool is based on logistic regression models that predict short-term and long-term job prospects based on, among other variables, age, gender, citizenship and health impairment \citep{allhutter_algorithmic_2020}. Based on predictions from these models, job seekers are placed in one of three job prospects group.  According to \cite{allhutter_algorithmic_2020}, those with mediocre job prospects are the focus of PES' measures to increase re-employment chances. Those in the highest group receive less intensive support by the PES as they are assumed to resume employment even without intensive support and those in the lowest group are mostly referred to an external institution. Due to negative coefficients of the mentioned predictors in the short-term job prospects model, people of higher age, female gender, non-EU citizenship or people with health impairment, are predicted lower prospects of finding a job in the short term. However, these characteristics are usually considered \textit{protected attributes} under anti-discrimination legislation. The AMS algorithm may discriminate against these groups. As \citet[][p. 7]{allhutter_algorithmic_2020} put it, "previously discriminated or marginalized groups are more likely to be classified as part of group C [the low job prospects group], which in turn reinforces existing inequalities as more discriminated populations of job seekers are more likely to receive less support." However, alleged discrimination of this system is heavily debated in Austria \citep{Buchinger2019}.

\citet{desiere2021} investigate similar fairness aspects of the statistical profiling system used by the Flemish PES \textit{VDAB}. They document that job seekers belonging to historically disadvantaged groups such as migrants, disabled and older age groups are more often incorrectly classified as high risk of LTU (here, unemployment that last for more than six months). Although the statistical profiling approach is more accurate in predicting LTU than a simple rule-based approach, it also shows more discrimination (defined as the ratio of false positive rates between groups) towards the aforementioned groups. Even though sensitive characteristics are explicitly not included in the model, they are nonetheless present in the model as they correlate highly with predictors such as language skills. Moreover, \citeauthor{desiere2021} find a clear trade-off between accuracy (that is, the share of unemployed people correctly identified as LTU) and discrimination against minority groups. As accuracy increases, so does discrimination. At the same time, discrimination depends on the threshold used to determine whether someone is high risk or low risk. Increasing the probability threshold decreases discrimination as the share of minority groups decreases with higher thresholds. 

Investigating similar questions of fairness and their dependence on data analysis decisions is also the focus of this study. Before turning to our fairness evaluation, we first describe the data and prediction approach used in detail. 

\section{Methods}
\label{Methods}

\subsection{Data}
\label{Data}

Data are obtained from German administrative labor market records. These records are maintained by the Research Data Center of the German Federal Employment Agency at the Institute for Employment Research (IAB). Briefly speaking, they contain historic records of labor market activities (employment, unemployment, job search activities and benefit receipt) for large parts of the German population \citep[about 80\% of the German labor force][]{dorner2010}. Not included are self-employed individuals and civil servants as their data are processed by a different institution \citep{jacobebbinghaus2007german}. Overall, the data are of high quality because they are used by the German Statutory Pension Insurance to calculate pension claims and for assisting unemployed in finding a new job matching their specific profile \citep{jacobebbinghaus2007german}. 
In more detail, the data contain historic information from 1975 to 2017 on all individuals in Germany who meet at least one of the following conditions:  at least once in employment subject to social security (records start in 1975) or in marginal part-time employment (records start in 1999); received short-term unemployment benefits or participated in labor market measures in accordance with the German Social Code Book III (records start in 1975); received long-term benefits in accordance with the German Social Code Book II (records start in 2005); registered with the German PES as a job-seeker (records start in 1997); participated in an employment or training measure (records start in 2000) \citep{siab2019}. These information are recorded exact to the day and allow the creation of detailed individual labor market histories.

We use a 2\% random sample from the administrative labor market records, called \textit{Sample of Integrated Employment Biographies} \citep[SIAB,][]{siab2019}. They are \textit{integrated} as they combine information from various sources such as employment information, unemployment information and unemployment benefits receipt (see previous paragraph). Specifically, we use the factually anonymous version of the SIAB (SIAB-Regionalfile) – Version 7517 v1\footnote{Data access was provided via a Scientific Use File supplied by the Research Data Centre (FDZ) of the German Federal Employment Agency (BA) at the Institute for Employment Research (IAB).}. Some potentially sensitive information from the original SIAB file were removed in this version of the SIAB data to meet privacy regulations. Still, the data are well suited for the purpose of predicting LTU due to their enormous volume and granularity: they contain detailed employment histories of 1,827,903 individuals documented in a total of 62,340,521 rows of data. 

The SIAB dataset comes in a longitudinal form. That is, we often observe multiple entries per person. Each time a person enters a new relevant labor market status (e.g., registered as unemployed or started a job subject to social security), a new entry is created. On average, we observe more than 34 data points for each of the nearly two million individuals. Note that it also possible that we observe only one entry for an individual, for example, if she was employed without any interruptions by the same employer. Depending on the type (e.g., employment episode, unemployment episode or benefit receipt episode) of a data point, socio-demographic characteristics such as age, gender, education and occupation as well as information on the duration of the episode (e.g. duration of unemployment), information on income and industry (for employment episodes), information on  participation in PES' sponsored training measures (for training measures episodes), or information on job search activities (for unemployment episodes) are available.

We restrict the SIAB data to data points referring to the period between January 1, 2010 and December 31, 2016. We exclude data referring to periods \textit{prior to 2010} as German legislators introduced fundamental labor market reforms between 2002 and 2005, which resulted in major socio-cultural, but also institutional changes in German labor market policies and fundamentally changed the way how unemployed people were supported by the German PES. In addition, new types of data were added to the SIAB during that time that capture individuals' labor market behavior in response to the reforms. 

Data collected \textit{after 2016} are excluded because our objective is to predict unemployment that lasts for at least one year. Therefore, the last year of labor market histories available is needed to determine whether an individuals who became unemployed by the end of 2016 became long-term unemployed or not. While one could include unemployment periods that start after 2016 but ended before December 31, 2017, it would introduce inconsistencies as we would obtain only non-LTU episodes in 2017 but no LTU episodes due to the right censoring of the data in December 2017. Therefore, we consider only unemployment episodes that started before 2017.

In addition, we remove all individuals who never became unemployed during the period of observation. Since we predict LTU, individuals who were never either LTU or non-LTU would be completely irrelevant. These restrictions leave us with 303,724 unique individuals and 643,690 unemployment episodes.\footnote{Note that individuals can contribute more than one unemployment episode to our data as they may become unemployed more than once during the period of observation.}

\subsubsection{Definition of long-term unemployment}
\label{LTU}

We follow the definition of LTU employed by the German PES. According to the German Social Code Book III, article 18/1, individuals are long-term unemployed if they are continuously unemployed for more than one year. Participation in labor market measures for the unemployed as well as periods of sickness or interruptions for other reasons of up to six weeks do not count as interruptions of an unemployment period. 

LTU is therefore identified if a data point refers to an unemployment episode with a recorded length of more than one year.\footnote{Relevant episodes are those flagged as "job seeking while unemployed" and "job seeking while not unemployed" if "not unemployed" is caused by a parallel episode of participation in a PES labor market measure (German Social Code Book III, article 18 in combination with article 16).} If an unemployment episode's duration is less than one year, we define it as non-LTU. Unemployment periods are recorded in the administrative labor market data once an individual registers as unemployed with the PES. Therefore, they allow us to identify the exact date a person presents herself as unemployed to the PES. Moreover, as the records we use are historic, we also observe the end date of an unemployment episode, which allows us to obtain the exact duration of an unemployment period and therefore to identify LTU. Note that the end of an unemployment episode does not necessarily imply exit into employment. It is also possible that unemployed individuals simply no longer consult with the PES.

Using the definition from above, 97,599 (15.2\%) out of a total of 643,690 unemployment episodes identified in the data are LTU episodes. Regarding individuals, we find that 79,361 (26.1\%) out of the 303,724 individuals in our data who ever became unemployed during the period considered experienced LTU at least once. LTU episodes and the number of affected individuals by year are shown in Table \ref{tab_LTU_desc}. Overall, the annual risk rates of entering LTU as shown in Table \ref{tab_LTU_desc} roughly match official rates of entry into LTU reported by the German PES \citep{BA2019}.

\begin{table}[]
\caption{LTU episodes and affected individuals, by year}
\begin{tabular}{lcccc}
                 & \textbf{\begin{tabular}[c]{@{}c@{}}Unemployment \\ episodes\end{tabular}} & \textbf{\begin{tabular}[c]{@{}c@{}}LTU \\ episodes\end{tabular}} & \textbf{Individuals} & \textbf{\begin{tabular}[c]{@{}c@{}}Individuals \\ experiencing \\ at least one \\ LTU episode\end{tabular}} \\ \hline
\textbf{2010}    & 105,137                                                                   & 15,872 (15.1\%)                                                  & 91,405               & 15,872 (17.4\%)                                                                                             \\
\textbf{2011}    & 95,597                                                                    & 14,813 (15.5\%)                                                  & 83,177               & 14,813 (17.8\%)                                                                                             \\
\textbf{2012}    & 90,408                                                                    & 14,865 (16.4\%)                                                  & 79,154               & 14,865 (18.8\%)                                                                                             \\
\textbf{2013}    & 88,988                                                                    & 14,324 (16.1\%)                                                  & 78,527               & 14,324 (18.3\%)                                                                                             \\
\textbf{2014}    & 87,158                                                                    & 13,529 (15.5\%)                                                  & 76,747               & 13,529 (17.6\%)                                                                                             \\
\textbf{2015}    & 86,692                                                                    & 12,688 (14.6\%)                                                  & 76,187               & 12,688 (16.7\%)                                                                                             \\
\textbf{2016}    & 89,710                                                                    & 11,508 (12,8\%)                                                  & 78,373               & 11,508 (14.7\%)                                                                                             \\ \hline
\textbf{Total} & 643,690                                                                   & 97,599 (15.2\%)                                                  & 303,724              & 79,361 (26.1\%)                                                                                            
\end{tabular}
\centering
\label{tab_LTU_desc}
\end{table}

\subsubsection{Predictors}
\label{Predictors}

To predict an individual's risk of LTU when becoming unemployed, we require a dataset in a one-observation-per-unemployment-episode form. That is, we consider the risk of LTU separately for each unemployment episode found in our data. We do not consider LTU on a per-individual basis because individuals can become unemployed more than once. We prefer our per-unemployment-episode solution to a per-individual solution as a new profiling would be conducted each time an individual registers as unemployed with the PES.

Since the SIAB data come in a longitudinal form, we often observe multiple data points per individual prior to becoming unemployed. To reduce these multiple observations to one observation per unemployment episode and individual, we count, for example, the number of unemployment episodes an individual experienced in the past or the total duration of employment episodes. These predictors summarize individual \textit{labor market histories}. In addition, we create a series of predictors that inform us about the \textit{last job} held by a person, e.g., the industry branch of the job, the skill level required, and the (inflation-deflated) daily wage (if a person was ever employed). The choice of these predictors is inspired by other studies of statistical profiling cited in Section \ref{ProfilingOECD}, but also by our own work with the administrative labor market data \citep{bach2019}.

\textit{Socio-demographic} information is derived in two ways. Information such as age, gender, and German nationality are derived from the most recent data point containing such information observed prior to or at entry into an unemployment episode. For information such as education, we consider the highest value observed prior to or at entry into an unemployment episode as these characteristics are sometimes measured with some inconsistencies \citep{fitze2005}. 

In summary, our feature generation procedures ensure that only information observed at or before entry into unemployment is considered for predicting LTU. Table \ref{tab_pred_grouped} shows groups of predictors (socio-demographics, labor market history, last job) with examples for each group. Due to the large number of predictors (157), we list all predictors in the appendix only (Table \ref{tab_pred_all}). Table \ref{tab:soc-demo} shows the prevalence, for each year and by LTU status, of the socio-demographic groups that we will focus on in the fairness evaluation of the prediction models (female, non-German, non-German male, non-German female).  

\begin{table}[]\caption{Groups of predictors, with examples}
\begin{tabular}{ll}
\hline
\textbf{Group} & \textbf{Example predictors}   \\ 
\hline
\textit{Socio-demographics} & Age, State of residence, Education, Number of moves \\
\textit{Labor market history} & Total duration of unemployment episodes  \\
 & Mean duration of employment episodes  \\
 & Total duration of job seeking episodes, scaled by age  \\
 & Industry worked in the most \\
 & Time since last employment episode  \\
\textit{Last job} & Industry\\
& Duration of employment \\
& Skill-level required for last job \\
& More than one job\\
& Part-time/Full-time/Marginal employment \\
& Fixed-term employment \\
& Inflation-deflated average daily wage \\
\hline
\end{tabular}
\centering
\label{tab_pred_grouped}
\end{table}

\subsection{Prediction setup}
\label{Setup}

We use the outlined variables to predict the risk of LTU for an individual unemployment episode. Specifically, the prediction task includes the following components:

\begin{itemize}
\item Set of \textbf{nonsensitive attributes} $X$. This set includes all predictors that are presented in  Section \ref{Predictors}.

\item \textbf{Protected attribute} $S$. Members of unprivileged group, $S = s^*$, members of privileged group, $S = s$. Following Germany's main anti-discrimination regulation, Article 3 of the \textit{Grundgesetz}, we consider gender and German nationality as protected attributes, with female and non-German individuals representing the unprivileged groups. We furthermore consider two (unprivileged) subgroups based on the intersection of both attributes: non-German females and non-German males. Note that the protected attributes are \textit{not used} as predictors when building the prediction models. 

\item \textbf{Observed outcome} $Y \in \{0,1\}$. True binary label of long-term unemployed ($Y = 1$) and not long-term unemployed ($Y = 0$), as outlined in Section \ref{LTU}.

\item \textbf{Risk score} $R \in [0, 1]$. Estimate of $Pr(Y=1 \mid X)$. The predicted risk of becoming long-term unemployed based on a given prediction model. 

\item \textbf{Prediction} $\hat{Y} \in \{0,1\}$. Binary prediction of becoming long-term unemployed ($\hat{Y} = 1$) and not becoming long-term unemployed ($\hat{Y} = 0$). Generally, we assume that individuals whose unemployment episodes are classified as LTU would be eligible for labor market support programs. The classification is based on the risk score $R$ and can be assigned along different \textit{classification policies}: 

\textbf{Policy 1a (P1a)}. Assign $\hat{Y} = 1$ to the top 10\% episodes with the highest predicted risk scores. The classification threshold $c_{10}$ is the $(0.1 \times n)$-th largest element of the risk score vector $\mathbf{r}$.

$$ \hat{Y}^{(h_a)} = 1 \ \text{if} \ R \geq c_{10}, \ \text{else} \ 0 $$

\textbf{Policy 1b (P1b)}. Assign $\hat{Y} = 1$ to the top 25\% episodes with the highest predicted risk scores. The classification threshold $c_{25}$ is the $(0.25 \times n)$-th largest element of the risk score vector $\mathbf{r}$.

$$ \hat{Y}^{(h_b)} = 1 \ \text{if} \ R \geq c_{25}, \ \text{else} \ 0 $$

\textbf{Policy 2 (P2)}. Assign $\hat{Y} = 1$ to the 50\% of episodes with medium predicted risk scores. The classification threshold $c_{75}$ is the $(0.25 \times n)$-th smallest element of the risk score vector $\mathbf{r}$.

$$ \hat{Y}^{(m)} = 1 \ \text{if} \ c_{25} \geq R \geq c_{75}, \ \text{else} \ 0 $$

\end{itemize}

Among the three classification policies, P1a and P1b align with the common rationale of classifying high risk episodes to the LTU class. As we assume that being predicted as LTU would eventually result in interventions, e.g. special support by PES in practice, P2 focuses on a scenario in which such interventions are targeted to medium risk cases. This scenario is inspired by the Austrian AMS example which, allegedly, focused support measures to job seekers with a medium risk of LTU (see  Section \ref{FairProfiling}).

\subsubsection{Prediction models}
\label{Models}

We consider four methods for building prediction models of LTU. In addition to regression approaches, as e.g. used in the Austrian case \citep{allhutter_algorithmic_2020}, we focus on prominent ensemble methods that are typically well-suited for prediction tasks with many features and potentially complex relationships (see also the prediction methods used in \citealt{de2018predicting}). Besides differences in flexibility, the methods also differ in the interpretability of the resulting models, which can be a decisive factor when algorithmic profiling of  job seekers is put to practice by PES. In summary, we compute predictions based on: 

\begin{itemize}
\item \textbf{Logistic Regression (LR)}. Common (unpenalized) logistic regression, only main effects for all predictors are included. Results in an interpretable set of coefficients, and is included as a benchmark.

\item \textbf{Penalized Logistic Regression (PLR)}. Logistic regression with a penalty on the ($\ell_1, \ell_2$) norm of the regression coefficients \citep{Tibshirani1996}. In the former case ($\ell_1$ penalty), a more parsimonious model compared to unpenalized logistic regression can be returned, which may increase both interpretability and prediction performance. 

\item \textbf{Random Forest (RF)}. Ensemble of deep (uncorrelated) decision trees grown on bootstrap samples \citep{Breiman2001}. Results in a model that cannot be readily interpreted without further helper methods. 

\item \textbf{Gradient Boosting Machines (GBM)}. Ensemble of small decision trees that are grown in sequence by using the (updated) pseudo-residuals in each iteration as the outcome \citep{Friedman2000, Friedman2001}. Similar to RF, additional techniques are typically needed to support the interpretation of results.
\end{itemize}

\textbf{Model training and evaluation}. As outlined in Section \ref{Data}, our SIAB data includes information from the beginning of 2010 up to the end of 2016. To mimic a realistic profiling task, we build prediction models with \textit{training data} covering the years 2010--2015, and we compare and evaluate the resulting models with \textit{evaluation data} from 2016. To ease computational burden related to model tuning (see below), a random sample of 20,000 unemployment episodes from each training year (2010--2015) is drawn to construct the training set. Final model evaluation is done on the full data from 2016 (89,710 episodes).

\textbf{Model tuning and selection}. With the exception of unpenalized logistic regression, we need to tune hyperparameters to build a well performing model for a given task. The hyperparameter settings considered for each method are listed in Table \ref{tab:grids}. The respective best setting for each method is selected based on temporal cross-validation \citep{Hyndman2018}: Training and test sets are constructed from the 2010--2015 data by successively moving the time point which separates the fit and test period forward in time. While this leads the training data to grow over time, we fix the respective test period to a single year. That is, the first fit and test periods include data from 2010 (fit) and 2011 (test). The last fit period includes data from 2010--2014, and the last test period data from 2015. These data sets are used to repeatedly train and test models for each method-hyperparameter combination. 

\textbf{Training histories}. The hyperparameter setting that results in the best average test performance over time for each method is selected to re-train the respective final model with the \textit{full training data} (2010--2015). Furthermore, we re-train a second set of models with \textit{restricted training data} using only the year 2015. This is done to explore the performance and fairness implications of training LTU models with different training data histories. With respect to prediction performance, one might naturally expect limited performance of the second set of models as they have access to fewer training examples. However, another perspective is that the structural associations between the predictors and the outcome could change over time ("concept drift"), such that training only with newer data could be beneficial. With respect to fairness, one may argue that the second set of models has fewer chances to learn discriminatory practices with respect to the effects of gender and nationality on LTU propensities if those practices are more commonly observed in older (training) data. 

\textbf{Software}. We used Stata (15, \citealt{stata15}) and R (3.6.3, \citealt{RCoreTeam2020}) for data preparations. Model training and evaluation was done with Python (3.6.4), using the scikit-learn (0.19.1, \citealt{Pedregosa2011}) and aif360 (0.4.0, \citealt{aif360}) packages.

\subsubsection{Performance and fairness metrics}
\label{Metrics}

\textbf{Performance metrics.} A key aspect when considering statistical profiling of job seekers is whether the underlying prediction models can accurately identify individuals that face high LTU risks. This perspective considers accurate predictions as a prerequisite for optimal allocation of support programs to unemployed individuals, both overall and for members of protected groups. Prediction performance may be evaluated based on the predicted classes $\hat{Y}$ (accuracy, precision, recall, and F1 score) or based on risk scores $R$ (ROC-AUC and PR-AUC). 

\begin{itemize}
\item \textbf{Accuracy}. Classification accuracy of predictions compared to observed outcomes. In range $[0, 1]$.

$$ \text{Acc} = \frac{1}{n} \sum_{i=1}^n \mathbf{1} ( \hat{y}_i = y_i ) $$

Where we sum over $i = 1, \dots, n$ instances and $\mathbf{1}$ represents the indicator function.

\item \textbf{Precision (at k)}. Proportion of correctly identified LTU episodes among all \textit{predicted} LTU episodes. In range $[0, 1]$.

$$ \text{Prec} = \frac{1}{k} \sum_{i=1}^n y_i \mathbf{1} (r_i \geq r_{[k]}) $$

Where $k$ is a constant (i.e., the number of instances with a predicted positive outcome) and $r_{[k]}$ denotes the $k$-th largest element of the risk score vector $\mathbf{r}$.

\item \textbf{Recall (at k)}. Proportion of correctly identified LTU episodes among all LTU episodes. In range $[0, 1]$.

$$ \text{Rec} = \frac{1}{\sum_{i=1}^n y_i} \sum_{i=1}^n y_i \mathbf{1} (r_i \geq r_{[k]}) $$

\item \textbf{F1 Score}. Weighted average of precision and recall. In range $[0, 1]$.

$$ \text{F1} = 2 \times \frac{\text{Prec} \times \text{Rec}}{\text{Prec} + \text{Rec}} $$

\item \textbf{ROC-AUC}. Area under the receiver operating characteristic (ROC) curve. In range $[0, 1]$, with 0.5 representing a random model.

\item \textbf{PR-AUC}. Area under the precision-recall curve. In range $[0, 1]$.
\end{itemize}

\textbf{Fairness metrics.} We consider two perspectives on fairness. First, unemployed individuals that are members of unprivileged groups should not be disproportionately excluded from treatment. This (group-level) perspective considers support programs of PES as assistive interventions to which access should not be blocked or delayed just by virtue of being a member of a group that is defined by a protected attribute. Second, unemployed individuals with similar (nonsensitive) attributes should be assigned similar predictions. This (individual-level) perspective requires predictions that eventually make similar unemployed individuals equally eligible to be assigned to support programs. In both cases, we focus on fairness metrics that are defined solely based on predictions of long-term unemployment (i.e., no error-based metrics). This allows us to consider different policies, such as allocating support programs to either high or middle risk individuals (see  Section \ref{Setup}).

\begin{itemize}
\item \textbf{Statistical Parity Difference}. Difference in the probability of being predicted LTU -- i.e. being eligible for support programs -- between privileged and unprivileged groups.

$$ Pr(\hat{Y} = 1 \mid S = s^*) - Pr(\hat{Y} = 1 \mid S = s) $$

\item \textbf{Conditional Statistical Parity Difference}. Difference in the probability of being predicted LTU between privileged and unprivileged groups conditional on nonsensitive attributes. We condition on education (i.e., high education status).

$$ Pr(\hat{Y} = 1 \mid S = s^*, X = x) - Pr(\hat{Y} = 1 \mid S = s, X = x) $$

\item \textbf{Consistency}. Average similarity of individual predictions and the predictions of their k-nearest neighbors \citep{zemel13}. The neighbors are defined based on the full set of (nonsensitive) attributes. We use $n_{neighbors} = 5$. Higher scores indicate more consistent predictions.
\end{itemize}

$$ 1 - \frac{1}{n}\sum_{i=1}^n | \hat{y}_i - \frac{1}{n_{neighbors}} \sum_{j \in \mathcal{N}_{n_{neighbors}}(x_i)} \hat{y}_j | $$

\section{Results}
\label{Results}

\subsection{Performance comparison}
\label{Performance}

We present the prediction performance of the trained models in two steps. First, the model selection criterion and results from the temporal cross-validation procedure (with data from 2010--2015) are discussed to provide some context on the final prediction models that were chosen in this process. Second, we present the performance of the selected models in the evaluation set (data from 2016).

\subsubsection{Temporal cross-validation}

Model selection for PLR, RF, and GBM was done based on the average ROC-AUC over all test periods in the temporal cross-validation loop. Specifically, the respective hyperparameter setting with the highest average ROC-AUC was chosen for each method. Because we observe rather stable performance over time (i.e., no outlier years), strategies taking temporal performance variation into account  \citep{de2018predicting} are not necessary in our case. 

Table \ref{tab:train-perf}a shows the overall hold-out ranking performance (ROC-AUC) of the selected best models for each method over time. We observe ROC-AUC's in the range $[0.694, 0.774]$, which largely aligns with performance results that have been reported for LTU prediction in other countries (i.e., Belgium and New Zealand, \citealt{Desiere2019}). In comparison, we see that the logistic regression models are consistently outperformed by PLR, RF, and GBM. Given the difference between LR and PLR, we suspect that this is likely due to model specification issues (e.g., many correlated and potentially uninformative predictors) in the unpenalized, 'naive' logistic model. In addition, we observe a mildly positive trend for most models (except LR) with increasing test set ROC-AUC's over time, indicating that the models benefit from the increasing amount of historical training data that is used as we progress to more recent years. 

Classification performance of the selected best models based on threshold policy 1a is shown in Tables \ref{tab:train-perf}b and \ref{tab:train-perf}c. That is, precision and recall is computed based on class predictions in which unemployment episodes with risk scores that are within the top 10\% of all scores are classified as long-term unemployment episodes. Generally, the best results are achieved with PLR, RF, and GBM (except for the first test set, 2011), with GBM consistently showing the highest precision and recall scores. As with ROC-AUC, the recall scores of PLR, RF, and GBM tend to increase over time, while there is no clear trend of the corresponding precision scores.

\subsubsection{Evaluation set}

Ranking and classification performance metrics of the selected best prediction models (re-trained with data from 2010--2015) for the evaluation set (data from 2016) are listed in Table \ref{tab:test-perf}a. Starting with overall ranking performance, the findings confirm the temporal cross-validation results, with PLR, RF, and GBM outperforming the unpenalized logistic regression model. Among those three best approaches, we see little differences in ROC-AUC scores. However, the tree-based ensemble models improve over penalized logistic regression in terms of PR-AUC, indicating higher precision over the range of applicable classification thresholds. When following threshold policy 1a, i.e. thresholding at the top 10\% risk scores, the GBM model shows the best classification performance. Regarding precision, we see that among all unemployment episodes that are predicted to last longer than one year, 37.4\% are indeed long-term unemployment episodes. Conversely, 29.2\% of all observed LTU episodes in the 2016 data are detected and classified as such by the GBM model (recall). 

Complementing results for models that were trained using only data from 2015 are shown in Table \ref{tab:test-perf}b. Overall, it can be seen that there is a modest decrease in prediction performance in terms of ROC-AUC and PR-AUC when restricting the training data to only include the most recent year before the evaluation date cutoff. Although some improvements in precision and recall (at top 10\%) can be observed for LR and PLR, the GBM model still shows the best results both with respect to overall ranking and classification performance. 

\begin{figure}[!h]
\centering
\subfloat[LR]{\includegraphics[scale=0.3]{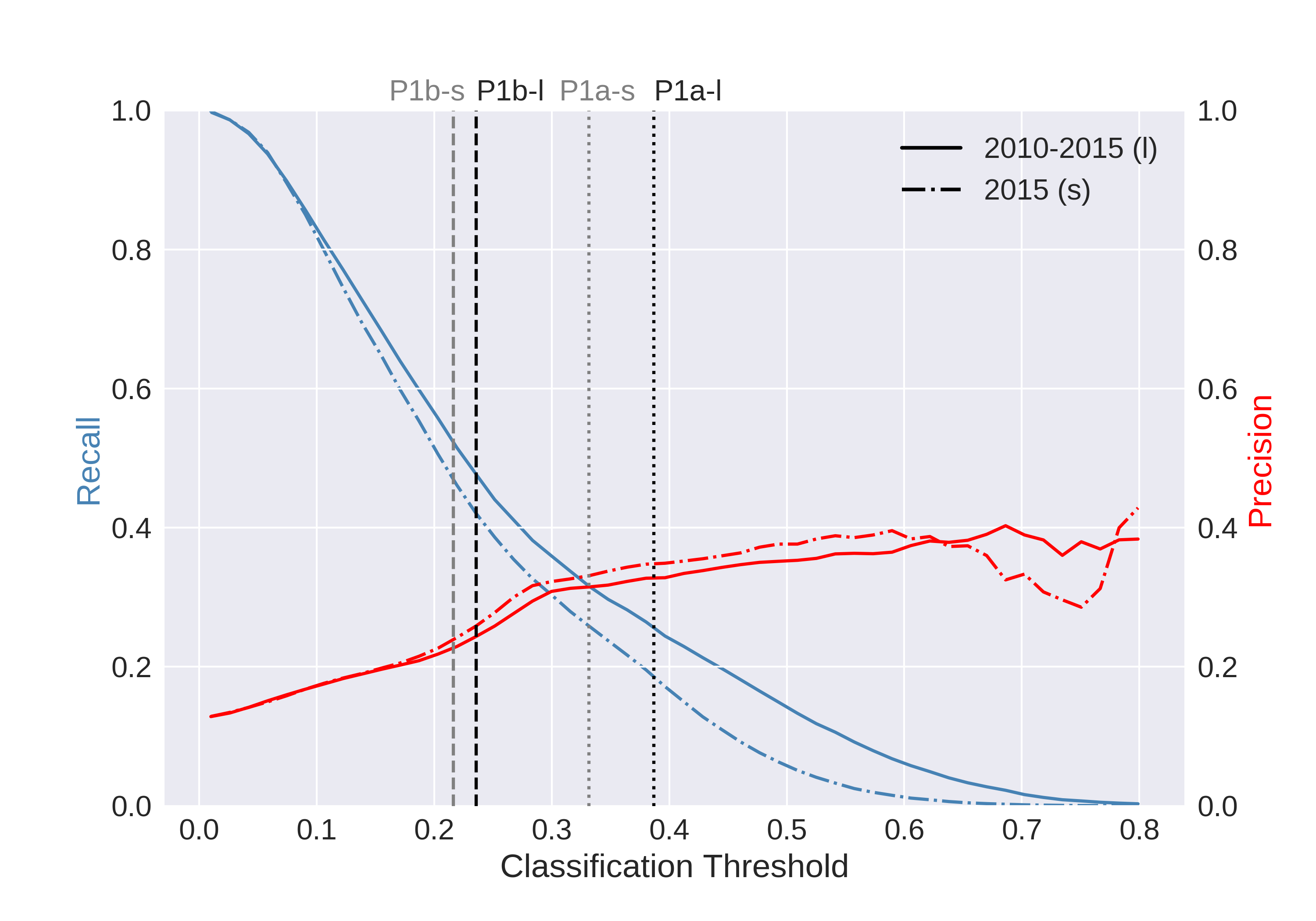}}%
\subfloat[PLR]{\includegraphics[scale=0.3]{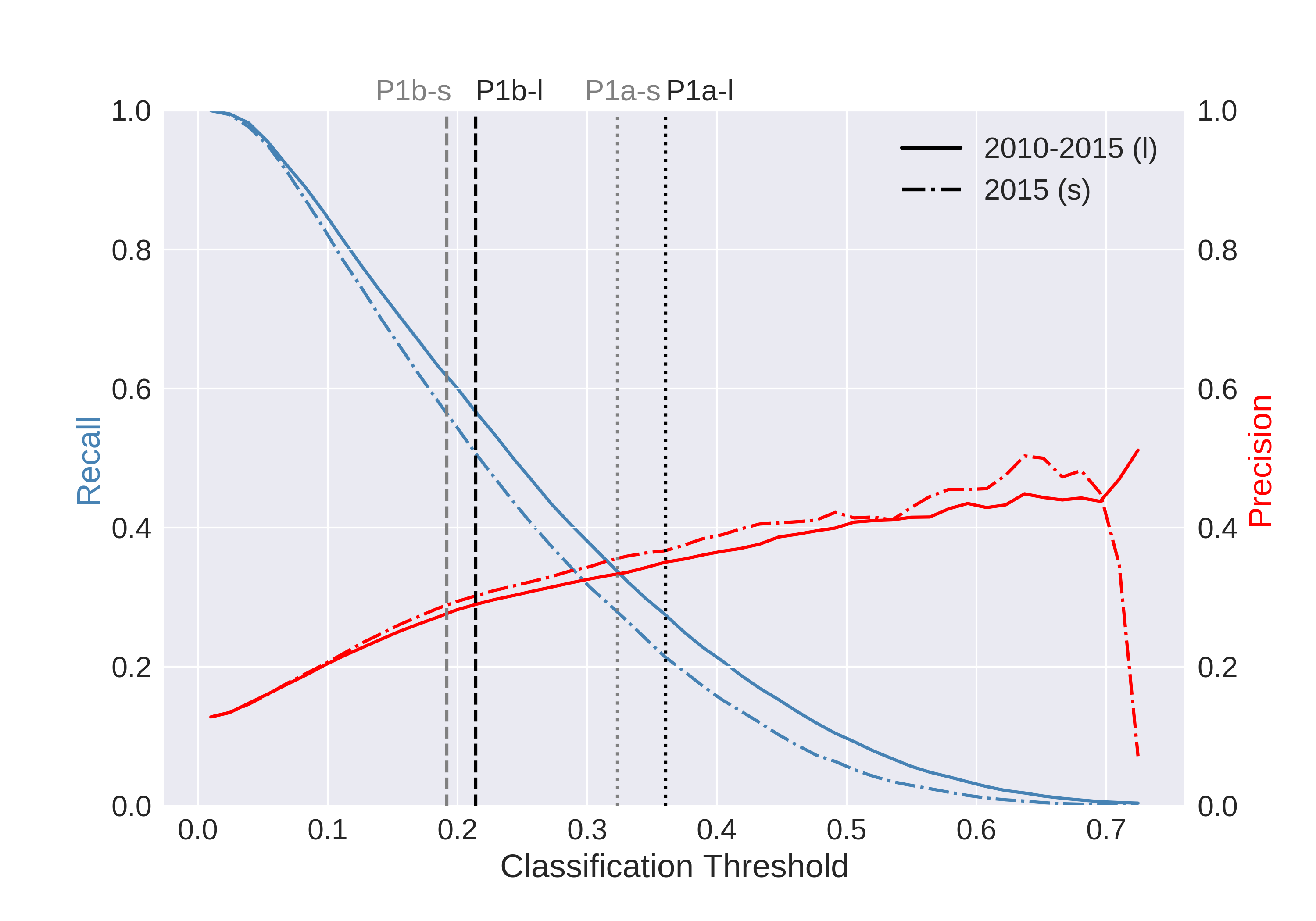}}\newline
\subfloat[RF]{\includegraphics[scale=0.3]{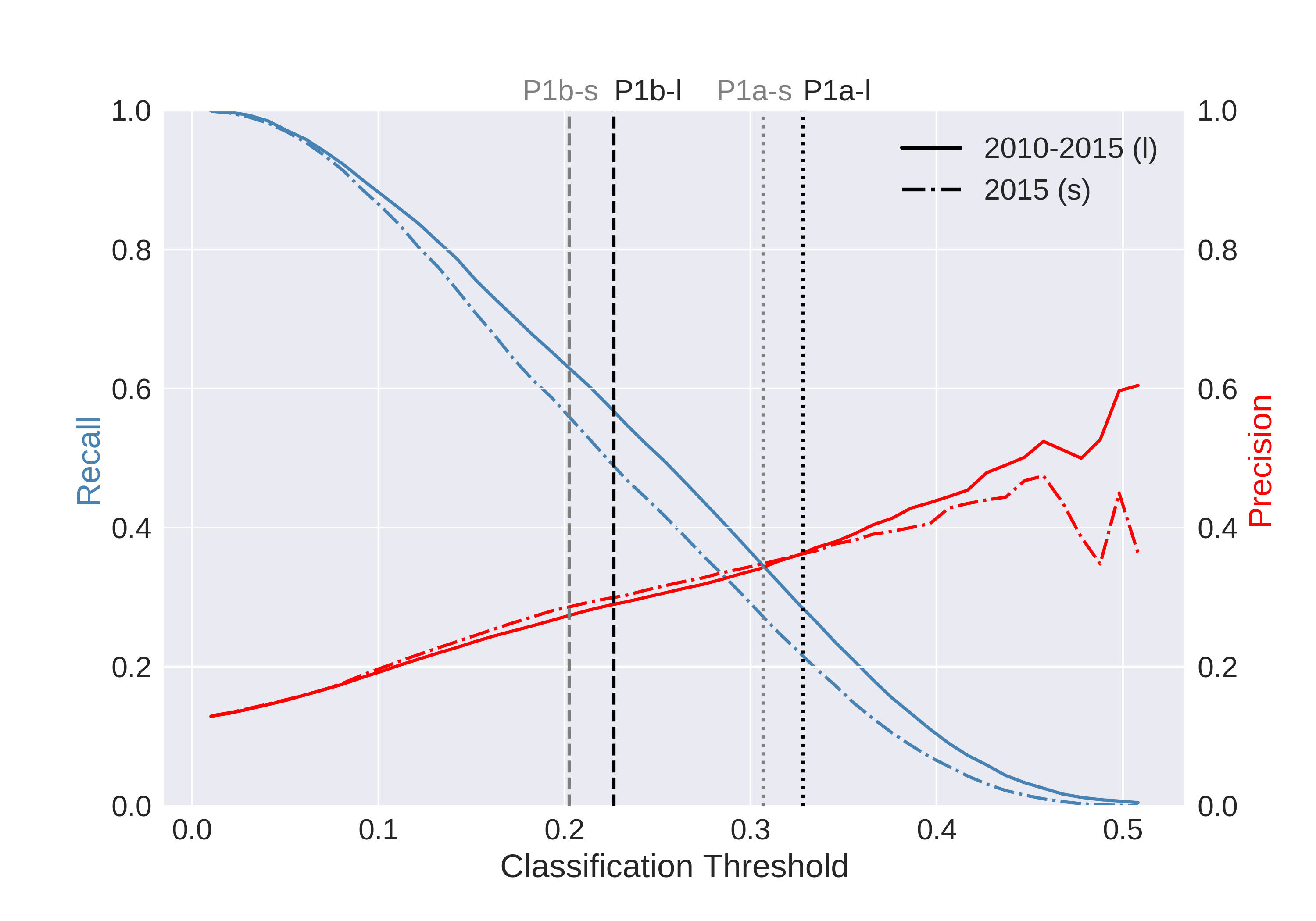}}%
\subfloat[GBM]{\includegraphics[scale=0.3]{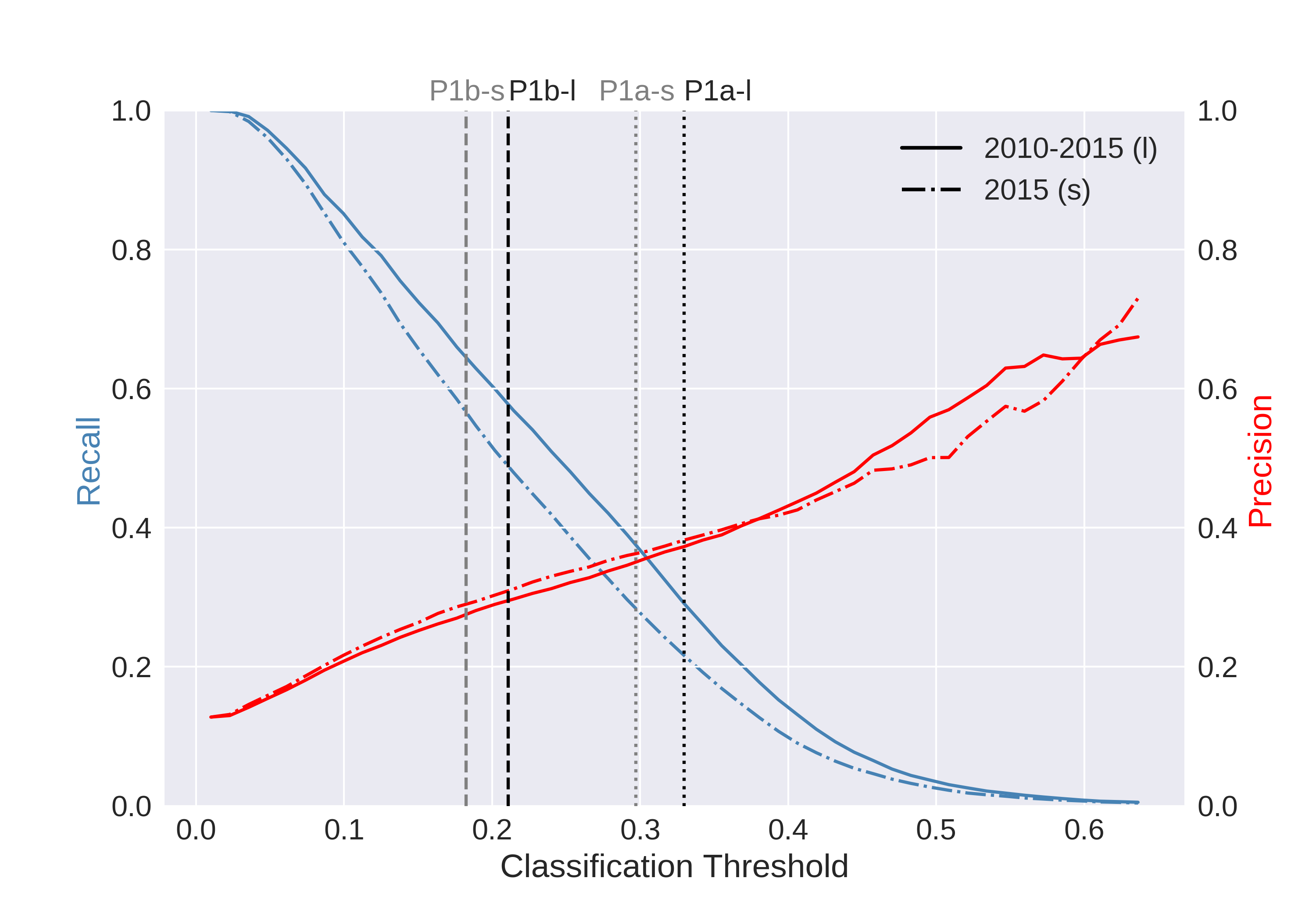}}%
\caption{Precision and recall versus threshold curves of selected prediction models (in 2016). The classification threshold of policy 1a is indicated by a dotted line, the threshold of policy 1b by a dashed line.}
\label{fig:pr-rec-curves}
\end{figure}

Precision and recall curves over the full range of applicable classification thresholds for all eight models (LR, PLR, RF, GBM trained with 2010-2015 and 2015 data, respectively) are presented in   Figure \ref{fig:pr-rec-curves}. The inherent trade-off between precision and recall is clearly illustrated: Moving from policy 1a (dotted lines) to policy 1b (dashed lines), i.e., less strict classification thresholds, increases recall at the cost of precision. That is, more true long-term unemployment episodes are detected but with lower precision. When using the full training data (2010--2015), PLR (0.565), RF (0.565) and GBM (0.576) achieve the best recall scores compared to LR (0.479) under policy 1b. At the same time, however, precision decreases (LR: 0.246, PLR: 0.290, RF: 0.290, GBM: 0.296). Thus, in this scenario we observe a higher number of false positives which, in practice, would result in more cases that would incorrectly be eligible for, e.g., job support programs. If we think of such programs as being rather expensive, the more strict policy 1a might be preferable (again, with the limitation that then more cases that should be eligible would be left out).  

\begin{table}[!h]
\centering
\caption{Prediction performance of selected prediction models, classification based on policy 1a (in 2016).}
\subfloat[Models trained with 2010-2015 data]{
\begin{tabular}{lrrrrrr}
\hline
& \textbf{ROC-AUC} &  \textbf{PR-AUC} &  \textbf{Accuracy} &  \textbf{F1 Score} &  \textbf{Precision} &  \textbf{Recall} \\
\hline
LR  &   0.700 &   0.256 &     0.837 &     0.287 &      0.328 &   0.256 \\
PLR &   0.760 &   0.298 &     0.842 &     0.308 &      0.351 &   0.274 \\
RF  &   0.764 &   0.312 &     0.845 &     0.320 &      0.365 &   0.285 \\
GBM &   0.770 &   0.325 &     0.847 &     0.328 &      0.374 &   0.292 \\
\hline
\end{tabular}}\newline
\subfloat[Models trained with 2015 data]{
\begin{tabular}{lrrrrrr}
\hline
& \textbf{ROC-AUC} &  \textbf{PR-AUC} &  \textbf{Accuracy} &  \textbf{F1 Score} &  \textbf{Precision} &  \textbf{Recall} \\
\hline
LR  &   0.695 &   0.253 &     0.838 &     0.291 &      0.332 &   0.259 \\
PLR &   0.756 &   0.298 &     0.843 &     0.312 &      0.356 &   0.278 \\
RF  &   0.757 &   0.296 &     0.842 &     0.306 &      0.349 &   0.272 \\
GBM &   0.763 &   0.310 &     0.844 &     0.318 &      0.363 &   0.283 \\
\hline
\end{tabular}}
\label{tab:test-perf}
\end{table}

\subsection{Fairness auditing}
\label{Fairness}

We next evaluate fairness aspects of the LTU predictions. We focus on the models trained with the full (2010--2015) and restricted (2015) training data, and audit their predictions in the 2016 data. We first base our fairness auditing on predicted classes that were assigned using fixed classification policies, before we move to more dynamic concepts. 

Table \ref{tab:test-fair} shows statistical parity differences, conditional statistical parity differences and consistency scores of LTU predictions that are based on the eight profiling models (LR, PLR, RF, GBM trained with 2010-2015 and 2015) and the three classification policies (P1a, P1b, P2), respectively. Parity differences were computed based on the attributes gender, non-German nationality, and for combinations of both attributes. For comparison purposes, these metrics were also computed for the true outcome ($Y$) as observed in the 2016 data. 

Starting with (unconditional) statistical parity, we see that the demographic composition of job seekers with long-term unemployment episodes is rather balanced with respect to gender and non-German nationality in the evaluation data. That is, we see slightly more LTU episodes among female (compared to male) and non-German female (compared to German and non-German male) job seekers, and fewer LTU episodes among non-German (compared to German) and non-German male (compared to German and non-German female) job seekers. Turning to statistical parity of the predictions, we observe that these 'true' differences are often elevated by the prediction models: Under policy 1a and 1b, (even) more unemployment episodes of female job seekers are classified as LTU, and (even) fewer episodes of non-German job seekers are commonly classified as LTU. An interesting pattern shows for policy 2. In this case, strong differences with respect to nationality emerge, with unemployment episodes of non-German job seekers now being much more likely to be treated as 'relevant' episodes ($\hat{Y} = 1$) that require attention. That is, non-German job seekers have a higher chance of being eligible for support programs under a policy that classifies unemployment episodes with medium LTU risk as relevant (P2), whereas German job seekers have a higher chance of being targeted under a policy that considers high LTU risks as relevant (P1a, P1b). This pattern holds for models that were trained with either the full or the restricted training data. 

Figure \ref{fig:risk-hist} shows that the observed differences when comparing policy 1a and 1b to policy 2 can be attributed to structural differences in the distributions of risk scores between both groups in terms of central tendency and skewness. Note that in all cases, neither gender nor non-German nationality were used as features when training the prediction models.

To complement the outlined findings, we present conditional statistical parity differences in Table \ref{tab:test-fair}. Following the argument that unemployment episodes of demographic groups may be more (or less) likely classified as LTU due to structural differences in nonsensitive attributes between those groups, we re-calculated statistical parity differences between groups conditional on having a high education status (i.e., a high school diploma). We see that for higher educated male and female job seekers, the proportions of unemployment episodes that are classified as LTU are rather similar. We also observe that the previously outlined differences with respect to nationality are mitigated to some extent. Nonetheless, even among higher skilled individuals, unemployment episodes of non-German job seekers are more often assigned to the 'relevant' class than those of German job seekers under policy 2. The opposite pattern still holds for most models under policy 1a and 1b, although on a lower level. 

In addition to group differences, we can also compare the class predictions with respect to consistency (Table \ref{tab:test-fair}). In this case, we are interested in evaluating whether  job seekers that are similar in nonsensitive attributes receive similar predictions on average. We observe rather high consistency scores for classifications that are based on policy 1a and 1b, indicating that job seekers with similar nonsensitive attributes would be largely treated similarly in these scenarios. Consistency is considerably lower under policy 2: Focusing on employment episodes with medium LTU risks decreases individual fairness such that predicted classes for similar job seekers are more heterogeneous. 

\begin{figure}[]
\centering
\subfloat[LR]{\includegraphics[scale=0.3]{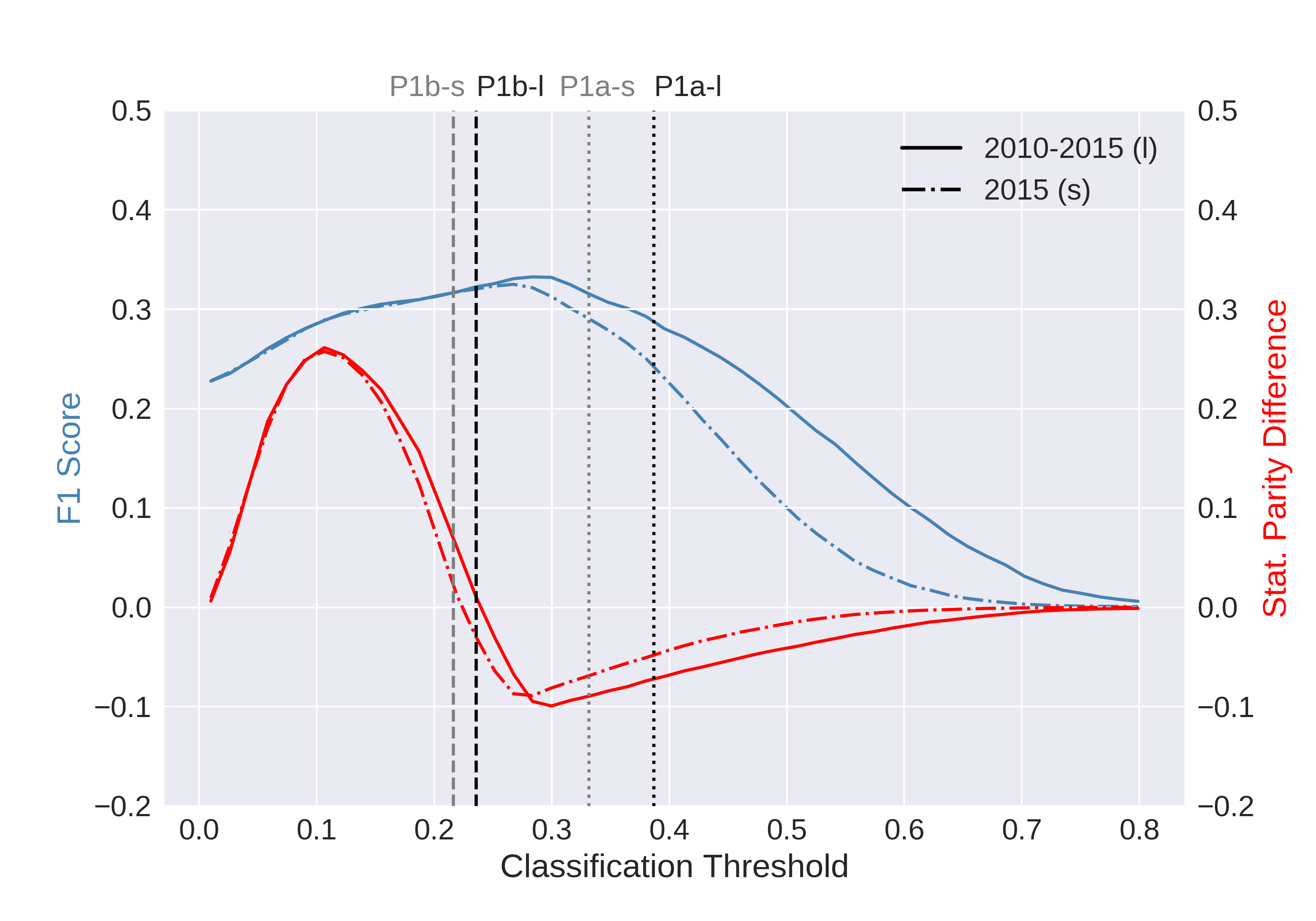}}%
\subfloat[PLR]{\includegraphics[scale=0.3]{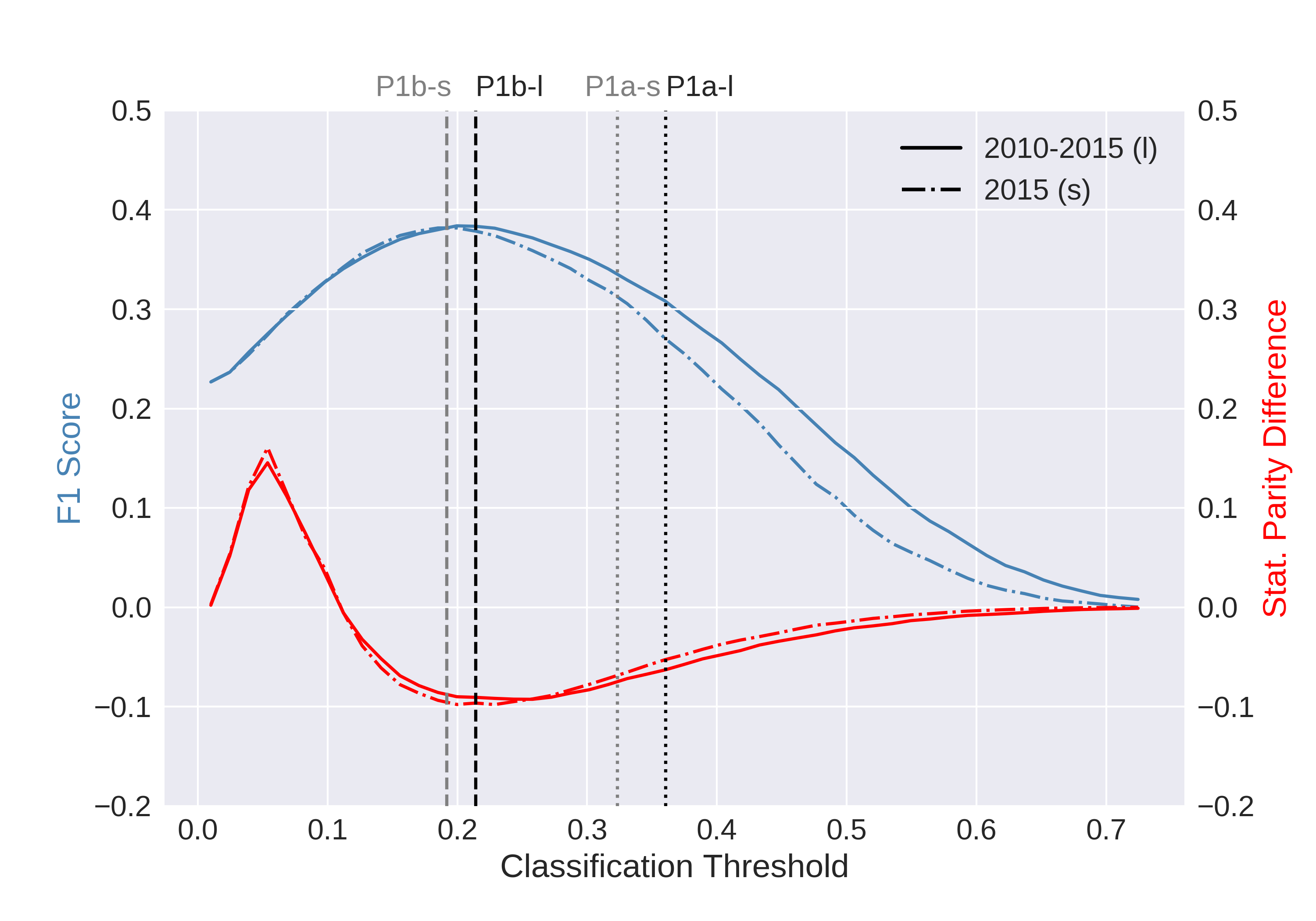}}\newline
\subfloat[RF]{\includegraphics[scale=0.3]{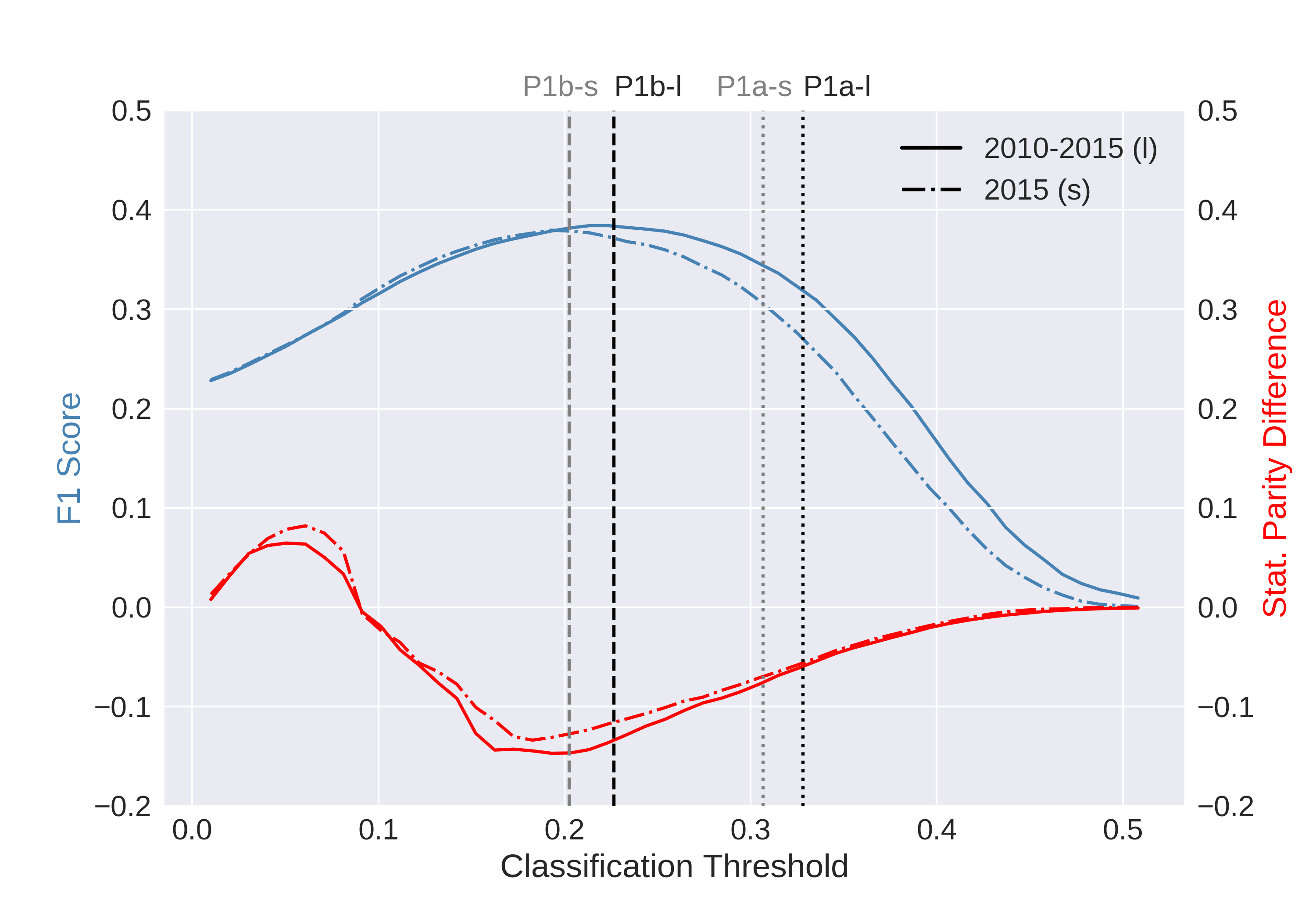}}%
\subfloat[GBM]{\includegraphics[scale=0.3]{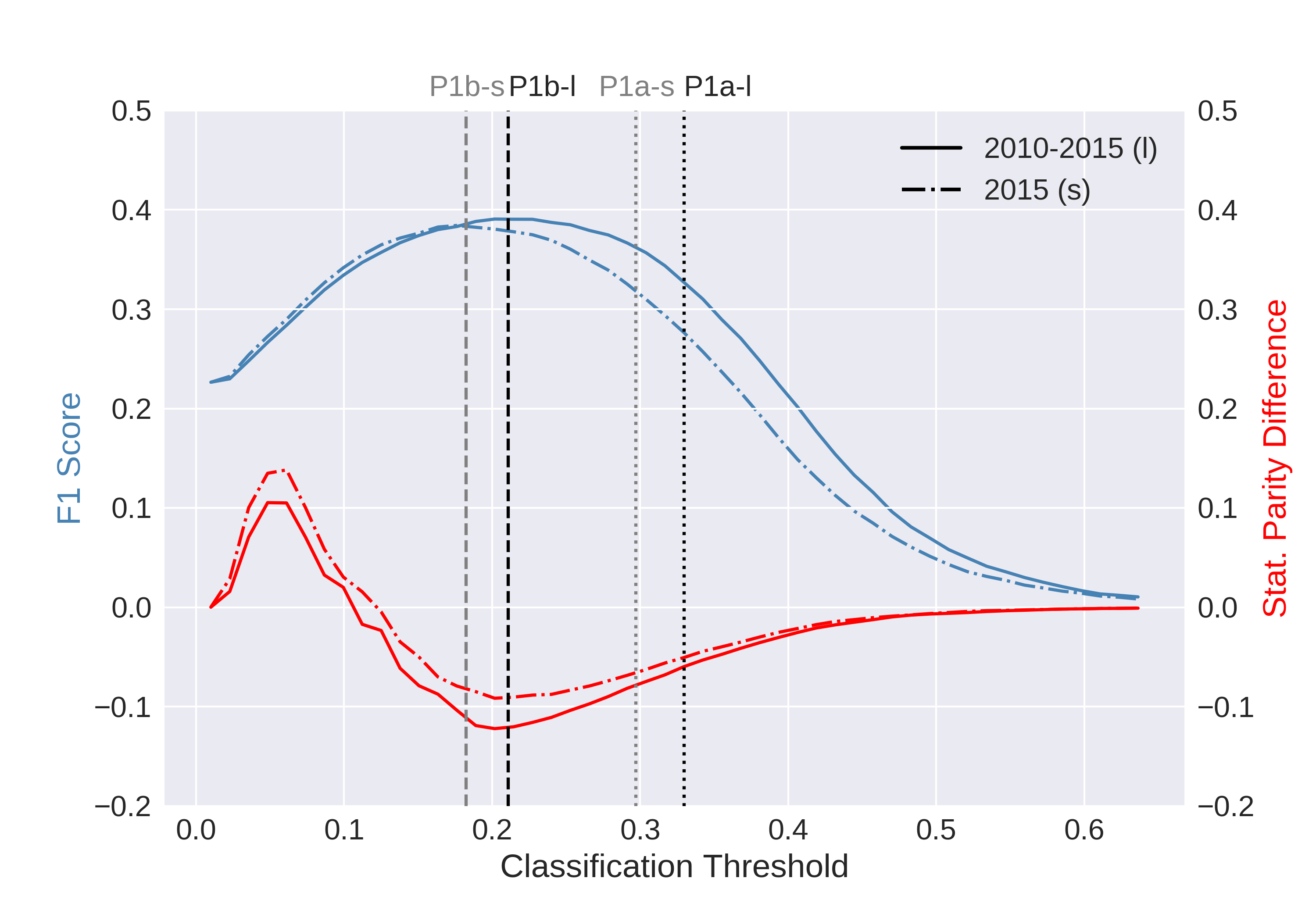}}%
\caption{F1 and statistical parity difference (non-German vs. German) versus threshold curves of selected prediction models (in 2016). The classification threshold of policy 1a is indicated by a dotted line, the threshold of policy 1b by a dashed line.}
\label{fig:fair-ger-curves}
\end{figure}

The fairness implications of different classification policies are further highlighted in   Figure \ref{fig:fair-ger-curves}. Here, we plot prediction performance (summarized by the F1 score) and statistical parity difference (based on nationality) over the full range of classification thresholds. For thresholds that are less strict than policy 1a and 1b, more unemployment episodes of non-German (compared to German) job seekers are classified as long-term unemployment, whereas this difference is reversed as we increase the classification threshold. For thresholds that are more strict than policy 1a and 1b, the composition of job seekers that would be targeted becomes more balanced.

\begin{sidewaystable}[!h]
\centering
\caption{Fairness evaluation of selected prediction models with different threshold policies (in 2016).}
\begin{tabular}{lll|rrrr|rrrr|r}
\hline
        &         & & \multicolumn{4}{c}{\textbf{Statistical Parity Difference}}    & \multicolumn{4}{c}{\textbf{Cond. Stat. Parity Diff.}}    &  \\
        &         & Training &         & Non- & Non-    & Non                    &         & Non- & Non-    & Non-                                                 & \textbf{Consis-} \\
Model & Policy    & data    &  Female & Ger. & Ger.\,M & Ger.\,F                &  Female & Ger. & Ger.\,M & Ger.\,F                                              & \textbf{tency} \\
\hline
\multicolumn{2}{l}{Observed $Y$} & &                   0.01 &                      -0.02 &                           -0.05 &                              0.02 &                        -0.02 &                            -0.02 &                                 -0.03 &                                    0.00 &         0.84 \\
\hline
LR & P1a & 2010-2015 &                   0.02 &                      -0.07 &                           -0.08 &                             -0.03 &                         0.00 &                            -0.03 &                                 -0.03 &                                   -0.01 &         0.96 \\
LR & P1a & 2015 &                   0.03 &                      -0.07 &                           -0.08 &                             -0.02 &                         0.00 &                            -0.03 &                                 -0.04 &                                   -0.01 &         0.96 \\
LR & P1b & 2010-2015 &                   0.03 &                       0.01 &                            0.00 &                              0.02 &                        -0.03 &                             0.07 &                                  0.09 &                                    0.02 &         0.92 \\
LR & P1b & 2015 &                   0.02 &                       0.02 &                            0.01 &                              0.02 &                        -0.04 &                             0.07 &                                  0.09 &                                    0.02 &         0.92 \\
LR & P2 & 2010-2015 &                   0.00 &                       0.20 &                            0.19 &                              0.17 &                        -0.00 &                             0.22 &                                  0.19 &                                    0.19 &         0.82 \\
LR & P2 & 2015 &                   0.01 &                       0.20 &                            0.18 &                              0.17 &                        -0.00 &                             0.18 &                                  0.15 &                                    0.17 &         0.82 \\
\hline
PLR & P1a & 2010-2015 &                   0.03 &                      -0.06 &                           -0.08 &                             -0.02 &                        -0.01 &                            -0.02 &                                 -0.03 &                                   -0.01 &         0.94 \\
PLR & P1a & 2015 &                   0.03 &                      -0.07 &                           -0.08 &                             -0.03 &                        -0.00 &                            -0.02 &                                 -0.03 &                                   -0.01 &         0.94 \\
PLR & P1b & 2010-2015 &                   0.06 &                      -0.09 &                           -0.14 &                              0.02 &                        -0.01 &                            -0.06 &                                 -0.08 &                                   -0.01 &         0.89 \\
PLR & P1b & 2015 &                   0.06 &                      -0.10 &                           -0.15 &                              0.01 &                        -0.01 &                            -0.06 &                                 -0.08 &                                   -0.01 &         0.89 \\
PLR & P2 & 2010-2015 &                  -0.02 &                       0.23 &                            0.25 &                              0.14 &                        -0.02 &                             0.17 &                                  0.18 &                                    0.10 &         0.76 \\
PLR & P2 & 2015 &                  -0.02 &                       0.26 &                            0.28 &                              0.14 &                        -0.02 &                             0.18 &                                  0.19 &                                    0.10 &         0.76 \\
\hline
RF & P1a & 2010-2015 &                   0.03 &                      -0.06 &                           -0.08 &                             -0.01 &                        -0.01 &                            -0.02 &                                 -0.04 &                                    0.00 &         0.94 \\
RF & P1a & 2015 &                   0.03 &                      -0.07 &                           -0.08 &                             -0.03 &                        -0.00 &                            -0.03 &                                 -0.04 &                                   -0.01 &         0.95 \\
RF & P1b & 2010-2015 &                   0.06 &                      -0.13 &                           -0.19 &                             -0.01 &                        -0.01 &                            -0.07 &                                 -0.10 &                                   -0.01 &         0.91 \\
RF & P1b & 2015 &                   0.06 &                      -0.13 &                           -0.18 &                             -0.01 &                        -0.01 &                            -0.07 &                                 -0.10 &                                   -0.01 &         0.91 \\
RF & P2 & 2010-2015 &                  -0.02 &                       0.19 &                            0.20 &                              0.12 &                        -0.01 &                             0.10 &                                  0.09 &                                    0.07 &         0.80 \\
RF & P2 & 2015 &                  -0.02 &                       0.21 &                            0.23 &                              0.11 &                        -0.03 &                             0.14 &                                  0.15 &                                    0.08 &         0.80 \\
\hline
GBM & P1a & 2010-2015 &                   0.03 &                      -0.06 &                           -0.08 &                             -0.01 &                        -0.01 &                            -0.03 &                                 -0.04 &                                   -0.00 &         0.93 \\
GBM & P1a & 2015 &                   0.02 &                      -0.07 &                           -0.08 &                             -0.03 &                        -0.01 &                            -0.04 &                                 -0.04 &                                   -0.02 &         0.92 \\
GBM & P1b & 2010-2015 &                   0.07 &                      -0.12 &                           -0.17 &                             -0.00 &                        -0.01 &                            -0.07 &                                 -0.10 &                                   -0.01 &         0.89 \\
GBM & P1b & 2015 &                   0.06 &                      -0.08 &                           -0.14 &                              0.03 &                        -0.01 &                            -0.06 &                                 -0.10 &                                    0.00 &         0.88 \\
GBM & P2 & 2010-2015 &                  -0.03 &                       0.23 &                            0.25 &                              0.13 &                        -0.04 &                             0.19 &                                  0.21 &                                    0.09 &         0.77 \\
GBM & P2 & 2015 &                  -0.04 &                       0.22 &                            0.26 &                              0.10 &                        -0.04 &                             0.19 &                                  0.22 &                                    0.08 &         0.76 \\
\hline
\end{tabular}
\label{tab:test-fair}
\end{sidewaystable}

\section{Discussion}
\label{Discussion}

We studied the use of statistical models for profiling job seekers based on German administrative data with respect to prediction performance and fairness implications. We compared regression and machine learning approaches using different classification policies and different training histories. On this basis, we particularly focused on the socio-demographic composition of the groups that would eventually be prioritized under different scenarios.

With respect to prediction performance, penalized logistic regression, random forests and gradient boosting outperformed unpenalized logistic regression, indicating that feature selection and/or regularization improved accuracy in our use case. The best model, based on gradient boosting, achieved a ROC-AUC of 0.77, which is competitive with approaches reported from other countries. It underlines that administrative labor market records can be used to build effective LTU prediction models. Nonetheless, using a top 10\% classification cutoff results in a recall score of 0.292 for the GBM model and thus leaves many true LTU cases undetected. Recall increases to 0.576 under a top 25\% cutoff policy for the same model, however, at the cost of a decrease in precision. Interestingly, using a restricted training data set that only includes records from the most recent year (before the temporal train-test split) has only a modest negative effect on performance. This is an interesting finding as it suggests that one could  retrain an effective model, for example, on a yearly basis to reflect only the most recent socio-cultural context and labor market conditions, a point also raised by \citet{desiere2021}. 

Furthermore, the results demonstrate that training a prediction model is only the first step. To implement a statistical profiling approach in reality, one needs to consider many design and policy decisions beyond pure technical details of the statistical prediction model. Decisions such as who should be targeted and how groups should be defined based on estimated risk scores have major consequences for the quality of the predictions. For example, determining the classification threshold cannot be done without consulting a country's socio-institutional context, for example, in terms of labor market policies, legislation and budget constraints. Here, we cannot make recommendations on which threshold should be used, but we highlight that different decisions imply different precision-recall trade-offs. We do not believe that policy makers need to be experts in training statistical models, but those who are should closely collaborate with them to reach informed decisions regarding the distribution of errors. For example, the question whether high precision may be preferred to high recall (or vice versa) can only be answered by policy experts and statistical experts together.

Awareness of the consequences of design decisions in profiling approaches becomes even more important when considering fairness aspects. Making uninformed decisions regarding the choice of the best classification threshold may result in inefficient PES spending when resources target many falsely classified people. However, if policy makers and those developing the profiling models are unaware of or ignore fairness audits, then the allocation of resources may not only be inefficient, but also discriminatory. For example, if PES support is based on risk groups derived from predictions of a statistical model that systematically assigns different risk scores to women than to men, PES risk discriminating based on gender. While some argue that such differences in scores simply represent the harsh reality of structural discrimination on the labor market, others argue that profiling systems will reinforce social inequality and foster disadvantages of already worse-off groups if they do not address fairness explicitly \citep[see, e.g.,][]{allhutter_algorithmic_2020}. At least, awareness of potential discrimination that may result from such systems is required. 

With regard to the fairness evaluations presented in this paper, we make four key observations. First, the models we trained tend to exacerbate differences between male and female and German and non-German job seekers with respect to their proportions of unemployment episodes that are classified as LTU, compared to the true differences between these groups. These differences were picked up by predictors that correlate with the protected attributes, as the latter were \textit{not} included in model training. Second, we observed stronger statistical parity differences for nationality than for gender. This result was consistently documented across prediction models and classification policies. Third, comparing non-German to German job seekers, different classification policies have considerably different consequences. Focusing on (very) high risk cases, (top 10\% or 25\% cutoff), unemployment episodes of non-German are less likely to be classified as LTU episodes, whereas the opposite holds true for a classification policy that focuses on the medium risk cases. Controlling for education (i.e., evaluating conditional statistical parity) mitigates, but does not equalize these differences. Fourth, although we expected that limiting the training data leaves fewer chances to learn differences between gender or nationality, we observed rather similar results when training with time-restricted data. Thus, our fairness evaluations indicate that discrimination may easily arise if one does not carefully assess whether LTU predictions differ across protected groups. 

If biased predictions are found, one may eventually want to correct them in many cases, for example, by pre-processing training data, by in-processing algorithms or by post-processing predictions \citep[see, e.g.][for an overview of fairness correction techniques]{caton2020fairness} such that subsequent decisions are free of discrimination. In some cases, however, biased predictions that result from structural discrimination may be explicitly accepted and become part of affirmative action policies. The German Social Code Book III, article 2/4, for example, states that PES support should explicitly improve the labor market chances of women to remove existing disadvantages. If one detects predictions biased against women, one could accept them and treat them as a means to reduce existing discrimination towards women. That is, structural differences in predictions may eventually be used in the decision-making step to mitigate historical discrimination \citep{kuppler2021distributive}. Nonetheless, detecting and understanding such differences is inevitable.

\section{Limitations} 
\label{Limitations}
There are several limitations to our study. We cannot evaluate how our results compare to current profiling approaches used by the German PES, also in terms of fairness evaluations. No information on the prediction performance or fairness of current profiling approaches in Germany are available. However, previous literature studying human predictions vs. predictions from simple rule-based and statistical models as well as the few studies that compared case worker-based and statistical profiling in other countries (see Sections \ref{Intro} and \ref{Background}) clearly show that statistical models outperform humans in various prediction tasks. The prediction performance of our statistical profiling approach is comparable to those of other countries \citep{Desiere2019}. For these reasons, we are confident that a similar conclusion regarding the superiority of statistical profiling to case worker-based profiling in terms of prediction performance would be reached for Germany if we were able to compare the approaches. In addition, we believe that our results provide only a lower bound in terms of prediction performance as we relied on an anonymized version of the administrative labor market data and because we did not include information such as regional unemployment rates and job vacancies in our models. Such data would likely be included if German PES were to use a statistical profiling approach similar to ours and it would likely increase prediction performance further.

\bibliography{references}

\appendix
\counterwithin{figure}{section}
\counterwithin{table}{section}

\section{Appendix}

\begingroup 
\small
\begin{center}
\begin{tabularx}{\textwidth}{l X}
\caption{List of predictors} \\
\hline
\textbf{Group} & \textbf{Predictor}   \\ 
\hline
\textit{Socio-} & Age     \\
\textit{demographics} & Education, categorized (6 dummy variables)     \\
& School education, categorized (7 dummy variables)     \\
& State of residence  \\
& Number of moves     \\
%Woman     \\
%German nationality    \\
%German nationality missing    \\
\hline
\textit{Labor market} & In Employment six weeks before unemployment?                               \\
\textit{history} & Long-term unemployment benefits receipt six weeks before unemployment?     \\
& Short-term unemployment benefits receipt six weeks before unemployment?    \\
& Subsidized employment six weeks before unemployment?                          \\
& Registered as job-seeking while not unemployed six weeks before unemployment? \\
& Registered with PES for other reasons six weeks before unemployment?          \\
& No information available six weeks before unemployment?                       \\
& Number of employers worked for                                                \\
& Number of jobs without any vocational training held                           \\
& Mean duration of employment without any vocational training                   \\
& Total duration worked in industry x (14 types of industries) \\
& Total duration more than one job                     \\
& Total duration in marginal employment                \\
& Total duration in full-time employment              \\
& Total duration in fixed-term employment             \\
& Total duration in temporary employment              \\
& Number of ALG II benefits receipt episodes          \\
& Total duration of ALG II benefits receipt episodes  \\
& Mean duration of ALG II benefits receipt episodes   \\
& Number of ALG I benefits receipt episodes           \\
& Total duration of ALG I benefits receipt episodes   \\
& Mean duration of ALG I benefits receipt episodes    \\
& Number of labor market program participation  episodes \\
& Total duration of labor market program participation  episodes   \\
& Mean duration of labor market program participation  episodes    \\
& Total duration of subsidized employment episodes    \\
& Number of job seeking episodes    \\
& Total duration of job seeking episodes   \\
& Mean duration of job seeking episodes   \\
& Industry individual worked in for the longest time (14 dummy variables)  \\
& Days since last employment, categorized (3 dummy variables)     \\
& Days since last labor market contact, categorized (4 dummy variables)     \\
& Days since last labor market contact (full-time), categorized (4 dummy variables)     \\
& Time since last unemployment spell, categorized (6 dummy variables)     \\
& Maximum of skill-level required for all employment episodes, categorized (4 dummy variables)   \\
& Total duration of employment episodes, scaled by age    \\
& Total duration of employment episodes with more than one job, scaled by age     \\
& Total duration of marginal employment, scaled by age     \\
& Total duration of full-time employment episodes, scaled by age     \\
& Total duration of fixed-term employment episodes, scaled by age     \\
& Total duration of temporary work episodes, scaled by age     \\
& Total duration of ALG II benefits receipt episodes, scaled by age    \\
& Total duration of ALG I benefits receipt episodes, scaled by age     \\
& Total duration of labor market program participation episodes, scaled by age    \\
& Total duration of labor market program participation (activation) episodes, scaled by age     \\
& Total duration of subsidized employment episodes, scaled by age     \\
& Total duration of job seeking episodes, scaled by age   \\
\hline
\textit{Last job} & No info about previous jobs available               \\
& Duration of last job                                \\
& More than one job at last job                       \\
& Inflation-deflated wage of last job                 \\
& Type of last job     \\
& Type of last job missing     \\
& Last job was part-time     \\
& Last job part-time missing     \\
& Skill-level required for last job, categorized (4 dummy variables)    \\
& Last job was fixed-term     \\
& Last job was fixed-term, missing     \\
& Last job was temporary work, missing        \\
& Last job was temporary work, missing         \\
& Industry of last job (14 dummy variables)  \\
& Commuted for last job?   \\
& Commuted for last job, missing    \\
& Last employment more than one job    \\
\hline
\end{tabularx}
\centering
\label{tab_pred_all}
\end{center}
\endgroup

\begin{table}[!h]
\centering
\caption{Prevalence of socio-demographic groups, by year and LTU}
\begin{tabular}{ll|rrrr|r}
\hline
              &                &                  & \textbf{Non-}  & \textbf{Non-}    & \textbf{Non-}    &  \\
\textbf{Year} & \textbf{Group} & \textbf{Female}  & \textbf{Ger.}  & \textbf{Ger.\,M} & \textbf{Ger.\,F} & \textbf{Obs.} \\
\hline
2010 & Overall  & 0.429 &     0.074 &          0.037 &            0.034 & 20,000 \\
     & No LTU   & 0.421 &     0.071 &          0.036 &            0.031 &  16,947 \\
     & LTU      & 0.476 &     0.089 &          0.037 &            0.047 &   3,053 \\
2011 & Overall  &  0.438 &      0.081 &        0.038 &             0.039 & 20,000  \\
     & No LTU   & 0.429 &     0.079 &          0.039 &            0.037 &  16,899 \\
     & LTU      & 0.486 &     0.093 &          0.037 &            0.050 &   3,101 \\
2012 & Overall  &  0.443 &      0.090 &        0.043 &             0.043 & 20,000  \\
     & No LTU   & 0.438 &     0.089 &          0.044 &            0.042 &  16,698 \\
     & LTU      & 0.471 &     0.096 &          0.042 &            0.050 &   3,302 \\
2013 & Overall  &  0.441 &      0.102 &        0.049 &             0.049 & 20,000  \\
     & No LTU   & 0.439 &     0.101 &          0.049 &            0.047 &  16,786 \\
     & LTU      & 0.447 &     0.108 &          0.045 &            0.059 &   3,214 \\
2014 & Overall  &  0.447 &      0.113 &        0.055 &             0.053 & 20,000  \\
     & No LTU   & 0.444 &     0.112 &          0.055 &            0.052 &  16,890 \\
     & LTU      & 0.467 &     0.118 &          0.055 &            0.058 &   3,110 \\
2015 & Overall  &  0.444 &      0.134 &        0.072 &             0.058 & 20,000  \\
     & No LTU   & 0.442 &     0.136 &          0.074 &            0.057 &  17,111 \\
     & LTU      & 0.460 &     0.126 &          0.056 &            0.065 &   2,889 \\
\hline
2016 & Overall  &  0.425 &      0.205 &        0.124 &             0.075 &  89,710 \\
     & No LTU   & 0.422 &     0.209 &          0.130 &            0.073 &  78,202 \\
     & LTU      & 0.445 &     0.177 &          0.084 &            0.087 &  11,508 \\
\hline
\end{tabular}
\label{tab:soc-demo}
\end{table}

\begin{table}[!h]
\centering
\caption{Tuning grids}
\begin{tabular}{l | l | l}
\hline
\textbf{Model type} & \textbf{Hyperparameter} & \textbf{Values} \\ 
\hline
Penalized Logistic & \texttt{penalty} & \texttt{l1, l2}  \\
Regression & \texttt{C} & \texttt{0.001, 0.01, 0.1, 1, 10, 100, 1000} \\
\hline
Random Forest &  \texttt{max\_features} & \texttt{sqrt, log2} \\
& \texttt{min\_samples\_leaf} & \texttt{1, 5, 10}  \\
& \texttt{n\_estimators} & \texttt{500, 750} \\
\hline
Gradient Boosting &  \texttt{max\_depth} & \texttt{3, 5, 7}  \\
Machines &  \texttt{max\_features} & \texttt{sqrt, log2}  \\
& \texttt{n\_estimators} & \texttt{250, 500, 750}  \\
& \texttt{learning\_rate} & \texttt{0.01, 0.025, 0.05}  \\
& \texttt{subsample} & \texttt{0.6, 0.8} \\
\hline
\multicolumn{3}{l}{\footnotesize Note: \texttt{scikit-learn} default settings are used for parameters not listed.}
\end{tabular}
\label{tab:grids}
\end{table}

\begin{table}[!h]
\centering
\caption{Temporal cross-validated prediction performance of selected prediction models}
\subfloat[ROC-AUC]{
\begin{tabular}{lrrrrr}
\hline
& \textbf{2011} & \textbf{2012} & \textbf{2013} & \textbf{2014} & \textbf{2015} \\
\hline
LR & 0.717 & 0.694 & 0.702 & 0.710 & 0.714 \\
PLR & 0.743 & 0.737 & 0.751 & 0.762 & 0.760 \\
RF & 0.764 & 0.744 & 0.754 & 0.765 & 0.763 \\
GBM & 0.768 & 0.749 & 0.763 & 0.774 & 0.772 \\
\hline
\end{tabular}} \\
\subfloat[Precision at top 10\% (policy 1a)]{
\begin{tabular}{lrrrrr}
\hline
& \textbf{2011} & \textbf{2012} & \textbf{2013} & \textbf{2014} & \textbf{2015} \\
\hline
LR & 0.370 & 0.370 & 0.385 & 0.370 & 0.350 \\
PLR & 0.327 & 0.401 & 0.403 & 0.395 & 0.377 \\
RF & 0.394 & 0.400 & 0.412 & 0.403 & 0.380 \\
GBM & 0.398 & 0.407 & 0.425 & 0.422 & 0.406 \\
\hline
\end{tabular}} \\
\subfloat[Recall at top 10\% (policy 1a)]{
\begin{tabular}{lrrrrr}
\hline
& \textbf{2011} & \textbf{2012} & \textbf{2013} & \textbf{2014} & \textbf{2015} \\
\hline
LR & 0.239 & 0.224 & 0.240 & 0.238 & 0.242 \\
PLR & 0.211 & 0.243 & 0.251 & 0.254 & 0.261 \\
RF & 0.254 & 0.242 & 0.256 & 0.259 & 0.263 \\
GBM & 0.257 & 0.247 & 0.264 & 0.272 & 0.281 \\
\hline
\end{tabular}}
\label{tab:train-perf}
\end{table}

\begin{figure}[!h]
\centering
\subfloat[LR]{\includegraphics[scale=0.4]{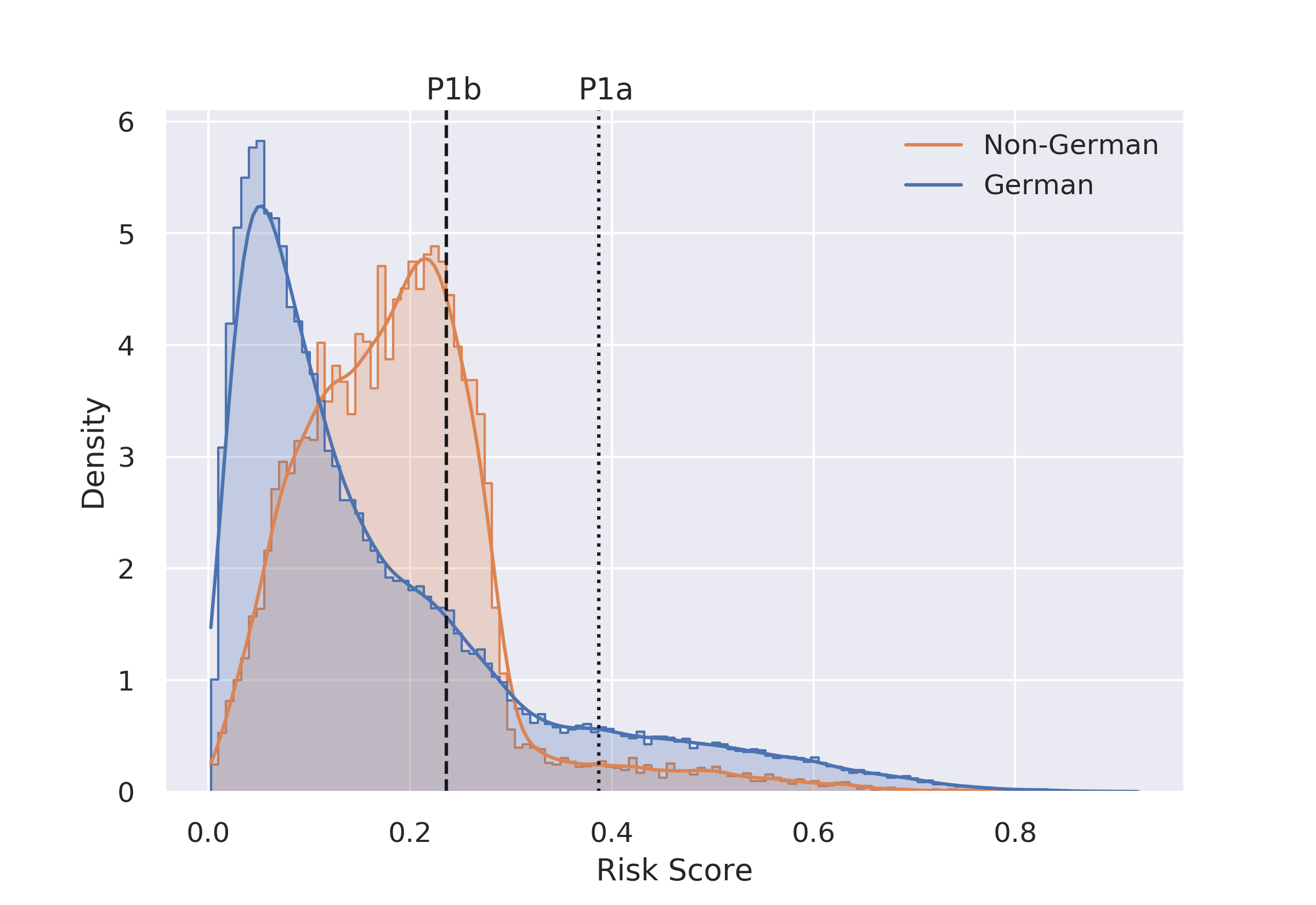}}%
\subfloat[PLR]{\includegraphics[scale=0.4]{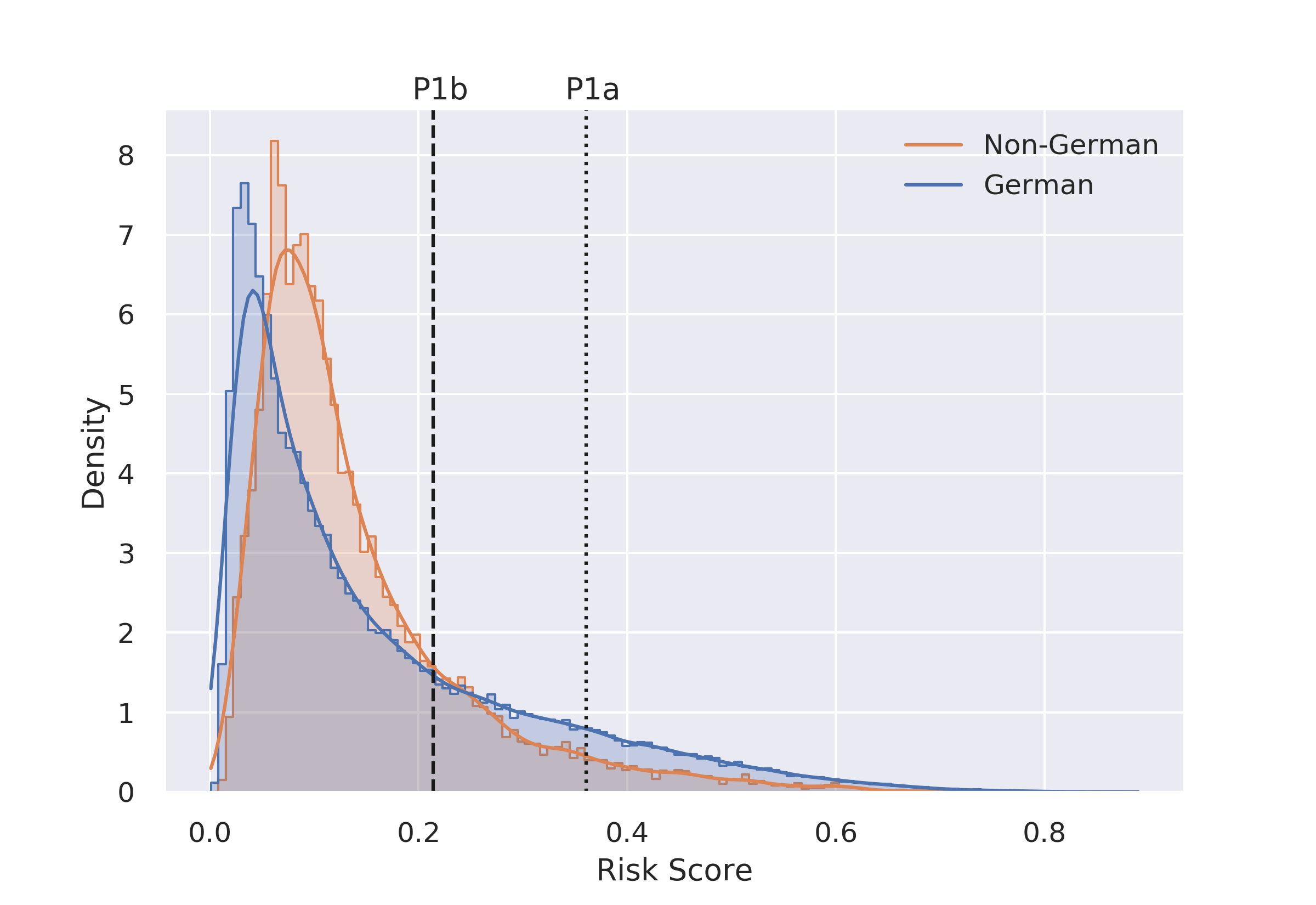}}\newline
\subfloat[RF]{\includegraphics[scale=0.4]{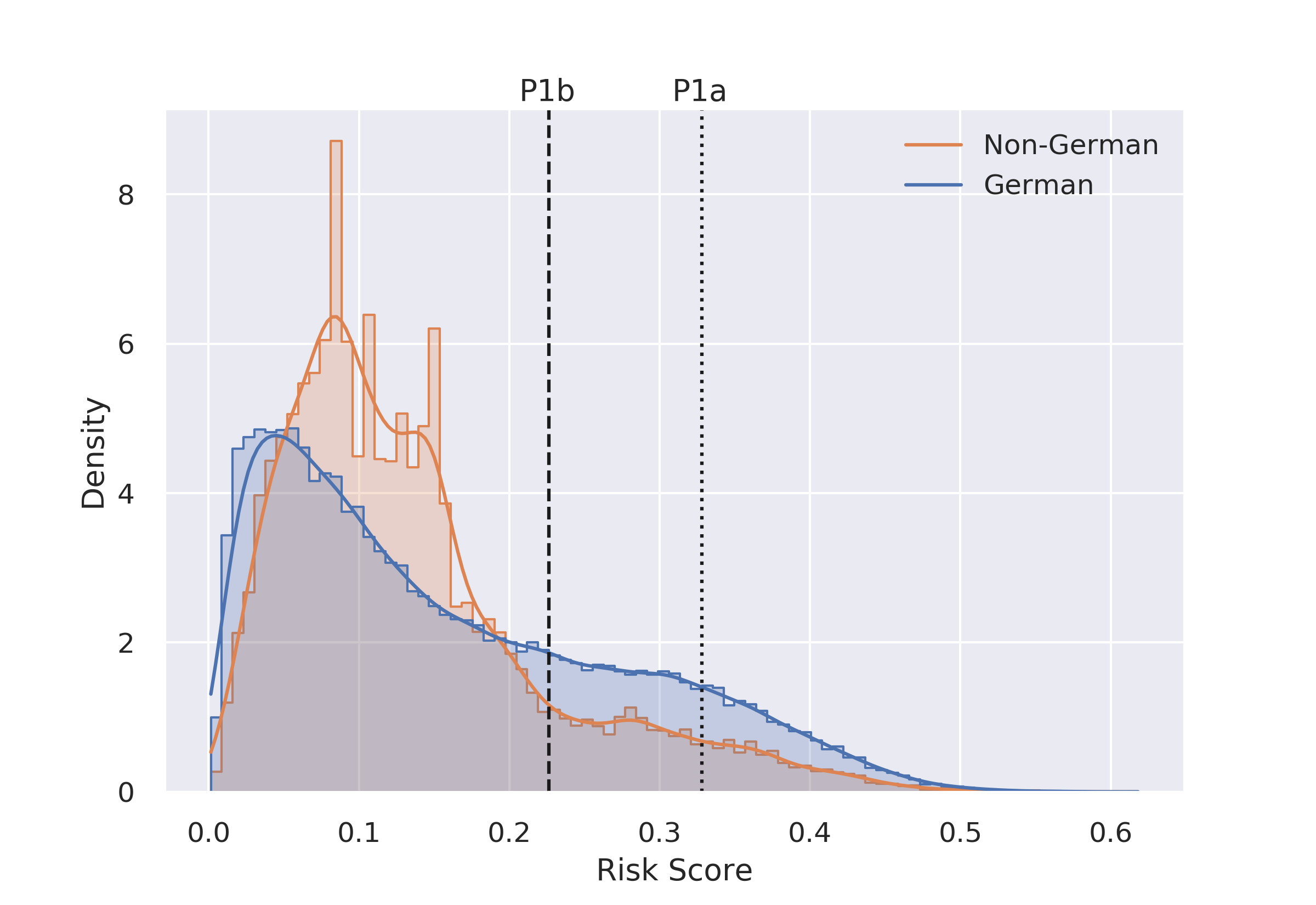}}%
\subfloat[GBM]{\includegraphics[scale=0.4]{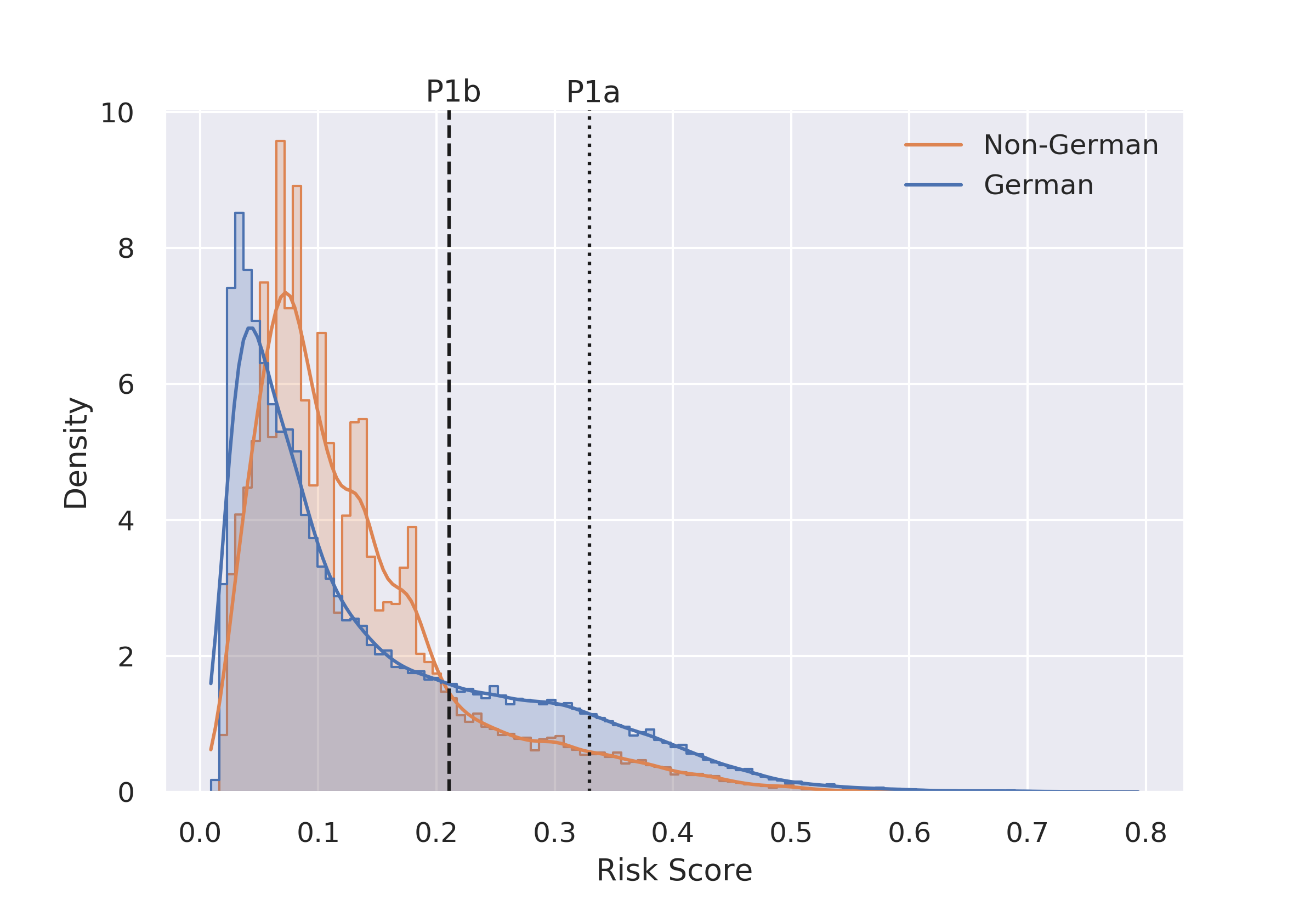}}%
\caption{Distribution of predicted risk scores (for German and Non-German) of selected prediction models (in 2016). The classification threshold of policy 1a is indicated by a dotted line, the threshold of policy 1b by a dashed line.}
\label{fig:risk-hist}
\end{figure}

\end{document}